\documentclass{emulateapj}
\usepackage{rotating}
\usepackage{epstopdf}

\slugcomment{submitted to AJ}
\shorttitle{AGN Discovered in the Kepler Mission}
\shortauthors{Shaya, Olling \& Mushotzky}
\begin{document}

\title{Active Galactic Nuclei Discovered in the Kepler Mission}
\author{Edward J. Shaya, Robert Olling, Richard Mushotzky}
\affil{Astronomy Department, University of Maryland,
    College Park, MD 20742}
\email{eshaya@umd.edu}
\begin{abstract} 
We report on candidate active galactic nuclei (AGN) discovered during the monitoring of $\sim$500 bright ($r < 18$ mag) galaxies over several years with the Kepler Mission.    
Most of the targets were sampled every 30 minutes nearly continuously for a year or more.  Variations of 0.001 mag and often less could be detected reliably.   
About 4.0\% (19) of our random sample continuously fluctuated with amplitudes increasing with longer timescales, but the majority are close to the limits of detectability with Kepler.  We discuss our techniques to mitigate the long term instrumental trends in Kepler light curves and our resulting structure function curves. The amplitudes of variability over four month periods, as seen in the structure functions and PSDs, can dramatically change for many of these AGN candidates.   Four of the candidates have features in their Structure Functions that may indicate quasi-periodic behavior, although other possibilities are discussed.
\end{abstract}


\section{Introduction}

A general characteristic of  emission from the centers of galaxies that contain active galactic nuclei (AGN) is large amplitude variability at all wavelengths from hours to years   \citep[e.g.,][]{Ulrich_etal1997,WebbMalkan_00}.
No other phenomenon in galaxies is sufficiently powerful to significantly alter the brightness of an entire galaxy on such long timescales.
Hence, surveys of galaxies designed to discover AGN based on variability \citep{Hook_etal94, Bershady_etal_98,
KlesmanSarajedini_07, Trevese_etal_08, MacLeod_etal10, Villforth_etal10, Sarajedini_etal11, Cicio_etal14} have been fairly successful.
Unfortunately, ground based time series usually suffer from interruptions due to weather, daylight, and telescope availabilities. 
Ground based studies also suffer from atmospheric seeing which limits photometric accuracy, typically to several $\times$ 0.01 mag, and can also result in false detections.
The Kepler satellite \citep{Borucki_etal10}, which can simultaneously monitor a large number of galaxies at a steady 30 minute cadence for months without disturbance from the atmosphere overcomes these limitations.  
 Detailed time series analysis benefits from these long continuous series.

Although the pixels in Kepler's CCDs subtend 4\arcsec\ on the sky, and hence the resolution is a factor 2-4 worse than typical ground based measurements, being in space compensates for this:  Kepler achieves photometric accuracy of better than 0.001 mag per 30 minute measurement at 17th mag and an order of magnitude better on timescales of $\sim$ day.   For an $L_*$ galaxy ($L \sim 2.6 \times 10^{10} L_\odot$, \citet{Schechter76}) with a typical central black hole of a few times $10^6 M_{\odot}$, the Eddington luminosity ($3.2 \times 10^4 M_{bh}/M_{\odot} L_{\odot}$) is about as bright as the galaxy.  AGN optical light output typically vary by large factors, and the amplitude of variability anti-correlates with AGN luminosity \citep{BarrMushotzky_86, Cristiani_etal96}.  Thus, Kepler data can reveal AGN with luminosities below $10^{-3}$ of their Eddington Limit, well below values reported previously.
However, this technique is fairly insensitive to Seyfert 2 type nuclei because the regions immediately around their BHs are not accessible in the optical.

We reduced data on $\sim 500$ anonymous galaxies that Kepler monitored during the period from Quarters 6 to 17, a duration of about 32 months.   
Only a few galaxies were monitored throughout the entire period,  but most have at least one year of data. 
In Quarter 17, Kepler suffered a second failure of its reaction wheels that ended the initial Kepler program.  However, the satellite has been resurrected as the new K2 Mission which relies on the two functioning reaction wheels, but is restricted to pointings in the ecliptic plane.

The optical variability of AGN is probably related to accretion disk instabilities, variations in accretion rates or changes in accretion disk structure.
The emitting regions are generally too small to be spatially resolved, therefore indirect means have been found to probe accretion disk dynamics and structure.
Analysis of variability leads to characteristic timescales that may be related to light travel times across emitting objects or to dynamical timescales such as orbital times.  
Typically, these timescales are set by the black hole mass and the accretion processes \citep{FrankKingRaine02}.
Rough estimates for some of  these timescales are: light-crossing $t_{l}= 2.6 M_7 R_{100}$ hr, dynamical $t_{dyn} = 10 M_7 R_{100}^{3/2}$ d and thermal cooling $t_{th} = 0.46 M_7 R_{100}^{3/2} \nu^{-1}$ hr,
where $M_7$ is the black hole mass in units of $10^7 M_{\sun}$, $R_{100}$ is the radius of the emission region in units of 100 times the Schwarzschild radius ($2GM/c^2$), and $\nu$ is the Shakura-Sunyaev viscosity parameter \citep{ShakuraSunyaev73}.
For typical AGN models, these timescales range from hours to years.
It is expected that, with sufficiently complete and accurate measurements, fits to the Power Spectral Density (PSD) function or Structure Function (SF) of an AGN's light curve might display features  at these key physical timescales.

Several teams have published results from Kepler data of known AGN 
\citep{CariniRyle12,ChenWang15,Revalski_etal14}.
\citet{Mushotzky_etal11} reported on four AGN, finding a power-law slope for the PSD with steep (-2.6 to -3.3) slopes.
\citet{Wehrle_etal13} have found PSD function fits by power laws with no clear breaks for four other AGN.  
The PSD of active galaxy Zw 229-15 was studied by \citet{Edelson_etal2014}, and was found to have a strong bend at the $\sim 5$ day timescale, which is similar to what has been reported in the x-ray PSD of AGN of similar mass.

Here, we try to make use of Kepler's exquisite photometric capabilities to explore nuclei with activity levels lower than would  normally be detectable and to focus on determining the true occurrence rates of AGN in galaxies.
In $\S 2$, we describe the target selection procedure and give some characteristics of the sample of galaxies observed.  
In $\S 3$, our photometry procedure is discussed.  The long term instrumental trends are removed from our aperture photometry by making use of the project supplied co-trending basis vectors.  In $\S 4$, a method to discriminate for activity is described and applied to find candidate AGN and structure functions are made from those light curves and analyzed.

\begin{figure}
\plotone{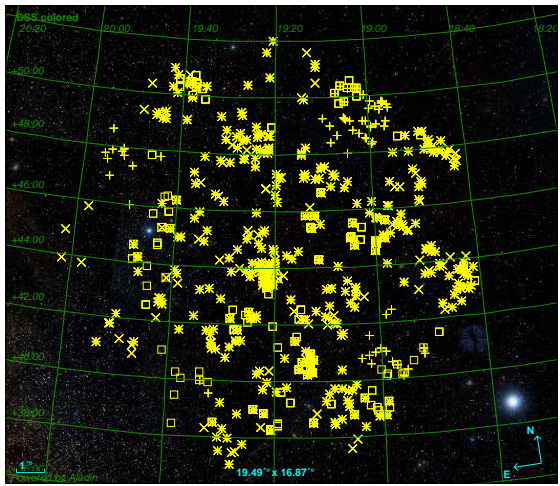}
\caption{Distribution of the galaxy sample on the sky in equatorial coordinates.  The symbols tell in which years the galaxies were observed.  Cycle 2, squares; Cycle 3, crosses; and Cycle 4, pluses.  The points are overlying a Palomar Sky Survey image of the Kepler Field.\label{skymap}}
\end{figure}

\section{Observations} \label{observations}
\subsection{Source Selection}

The Kepler Field, centered at  galactic $ l=264\fdg5,  b = +13\fdg5$, lies at fairly low galactic longitude and therefore had not been well covered by AGN surveys.  Pre-launch of Kepler, only a few AGN were known that resided in the field.  
Soon after launch, there were efforts to search surveys at various wavelengths to identify AGN based on colors
 \citep{EdelsonMalkan2012} or x-ray flux.
But, for our purpose, to discover the true occurrence rate of AGN, we simply selected a set of undistinguished
 galaxies of small angular extent, mostly from the 2MASS survey, to be monitored with Kepler.  
Because the data rate from the Kepler camera was so large and the download bandwidth was limited, specific
 targets in the Kepler Field were competitively awarded each year, and only a small patch of CCD pixels around
 each target image was downloaded.
Therefore the galaxies had to be small in angular extent to minimize the number of pixels for downloads.
Our target galaxies (in programs GO20058, GO30032, and GO40057) were extracted from the 2MASS extended source catalog (2MX; \citet{Jarret_etal00}).  
The target apparent Kepler magnitudes were restricted to $<$ 17.3 mag, and we selected small galaxies as defined by 
the 20th mag K-band isophotal radii, including $3 \sigma$ errors, 95\% were $<$ 2 pixels and the average was $3.7 \pm 1.3$ arcsec, or $0.93 \pm 0.33$ Kepler pixels.
Approximately every 3 months the spacecraft was rotated $90\degr$ to maintain sunlight on the solar panel and to download data.  This defines a Kepler Quarter.  
Targets were deselected if they fell between CCDs after any rotation.   According to the Explanatory
 Supplement (Cutri et al 2000), at these low galactic latitudes, about 80\% of the extended 2MX sources are
 galaxies and most of the rest of the sources are multiple stars.  
Our selection of 2MX galaxies contained only sources that were visually confirmed by the 2MX authors.
This procedure resulted in 1,858 galaxies as bright as 10th mag, and the median was near 15 mag.  
We then selected 400 galaxies from this set that were best separated  from other sources in the catalog and least contaminated from the overlapping scaled point spread functions of neighbors.  Their distribution on the sky is show in Fig.~\ref{skymap}.

\begin{figure}
\epsscale{1}
\plotone{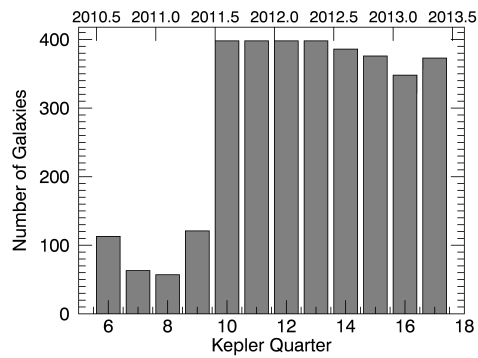}
\caption{The number of galaxies monitored per Kepler Quarter and assigned to this team's programs. \label{histn}}
\end{figure}

\begin{figure}
\epsscale{1.0}
\plotone{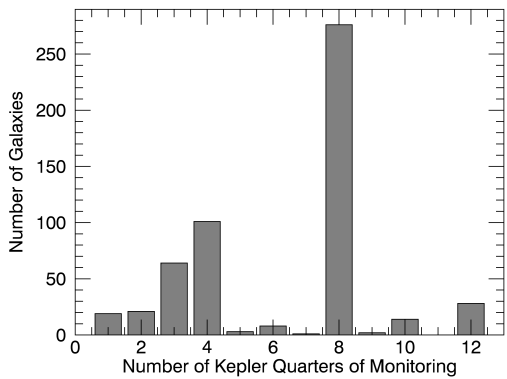}
\caption{Histogram showing how the target galaxies are distributed by the number of Kepler Quarters they were observed. \label{histquarters}}
\end{figure}

Our list of galaxies was proposed in Cycle 2, but only 130 were accepted for downloading and some for less than 4 Quarters.  
In Cycles 3 and 4 most of the galaxies on the list were selected for downloading.  
Figure~\ref{histn} shows the number of galaxies awarded to this project and actually monitored in each Quarter.  
For two years about 400 galaxies were monitored, and in total, 519 galaxies were assigned to be monitored.
Quarter 17 was cut short due to the failure of a reaction wheel and the photometry in that quarter appears to have suffered instabilities in most cases. 
Figure~\ref{histquarters} presents the number distribution of the number of Quarters observed.  
The distribution of the sample on the sky is shown in Figure~\ref{skymap}.

\section{Data Reduction}

The Kepler science data pipeline begins with a calibration (performed by the CAL module) of  the CCD images \citep{Quin10} by correcting for;  bias level, dark current (insignificant), smear due to shutterless readouts,  nonlinear gain (median of 112 $e^-$/ADU), certain undershooting pixels,  and individual pixel sensitivities (flat field).  
The pipeline also produces calibrated light curves in its Photometric Analysis (PA) package, based on the total counts above background in an "optimal aperture" around the target defined by the project.  
The optimal  aperture is a set of contiguous pixels within the downloaded stamp, fixed throughout a Quarter, with count rates from the target high compared to the background but avoids possible light contamination from nearby sources.   
Adjustments are made for cosmic rays and the observation time array is converted to barycentric times.  
Arrays of centroid positions and background light as functions of time are also created.   
The next stage of reduction, the Presearch Data Conditioning (PDC) \citep{Stumpe12}, applies corrections based on both known instrumental and spacecraft anomalies and correlated systematic photometric excursions found in the data.  
Excess flux in the source apertures due to unavoidable contamination from nearby stars is removed and outlier data points are flagged. 

Variations in PA counts are found to correlate with focus and PSF changes which arise from flexing of the telescope structure.   
Thermal transients, lasting several days, occur following downlinks, after safe modes, or   
from the changing orientation of the spacecraft with respect to the Sun.
Drift in pointing at the sub-pixel level also modifies the counts.  
Differential velocity aberration across the large field unavoidably causes the position of target images to slowly change, especially at the edges of the field.  
In the PDC process, a set of 14 normalized vectors, called cotrending basis vectors (CBV), are generated by singular value decomposition that represent correlated instrumental artifacts in the LCs in each CCD and in each Quarter.  
For the PDC light curves,  a superposition of these vectors that minimizes the rms deviations for a target over the Quarter is subtracted from the light curves.
    
It was recognized that the procedure of minimizing deviations could inadvertently remove astrophysical features in the time series and therefore, whenever the variations exceed some strength and are clearly intrinsic to the target, this step is either skipped or fewer CBVs are used.  
For  known AGN (in other GO data) with  large amplitude ($>5$\%)  variations, the PDC light curves supplied by the Kepler Project seem reasonable, and it appears that either no CBVs, or just the first one or two, are used.
However, in some of our target galaxies with lower amplitude variations,  the PDC procedure may have removed some of the long timescale physical fluctuations.
Usually, one can still discern that these are active galaxies because the rms in the light curves remains elevated compared to the very quiet galaxies, and power in the PDC is shifted from longer to shorter timescales.
Other examples of PDC over correction are the supernovae discovered in our programs (Olling et al 2014) in which the PDC light curves are so distorted that the supernova events were rendered nearly unrecognizable.  

Another problem is that many quiet galaxies have light curves that appear active in certain Quarters.  
 In a few cases, this occurs because the PDC procedure just failed. 
 In other cases, the CCD had undergone a period of excess noise arising from crosstalk between the CCDs and/or the 4 fine guidance sensors or from a high frequency amplifier oscillation as has been documented for Kepler CCDs \citep{Kolo10}.
In most of these cases it is necessary to simply disregard entire Quarters for targets on these CCDs to ensure reliable measurements.

\subsection{Generating Long Term Light Curves}
We reprocess our galaxy data into light curves starting from the Kepler pipeline CAL output of postage stamp cutouts with the background subtracted.  
By using more pixels than the PDC procedure, the long timescale trending effects from pointing drift and focus changes are reduced at the expense of some additional readout and sky noise.   
For our purposes, this tradeoff is highly beneficial because the former greatly exceeds the latter. 
Therefore, we sum counts in either $3 \times 3$ or $5 \times 5$ pixel apertures.  
With $5 \times 5$ pixels, the trends over a Quarter, which typically have an amplitude of 3 - 4\% in the PA light curves, drop to under 1\%.  
We do not use $7 \times 7$ pixel apertures because they often suffer significantly from contaminating variations of neighbors and from larger errors arising from a given background error.

We looked at two methods for correcting the long term trends in the counts due to instrumental effects.  
One method is to determine PSF widths and centroids in two dimensions and then find first order correction factors on the four parameters.
We could have used the Kepler pipeline determined widths and centroids, but it is easy to calculate simple stand-ins for these measures.  
For width changes we use the ratio of counts in the $5 \times 5$ pixel aperture to that in the $3 \times 3$ pixel aperture,  and for centroid shifts in each direction,  we use the difference in the counts in the pixels on either side of the center divided by the three contiguous central pixels.
Since the trends in counts have amplitudes of only a few percent, it is reasonable to assume that terms linear in these parameters would dominate. 
By assuming the coefficients are constant over each quarter, one can simultaneously solve for coefficients that minimize the variations.
For the quiet galaxies in our sample, this procedure typically reduces the trends to less than 0.1\% variability over 3 months.  
This confirms that much of the trending in Kepler photometry is due to focus changes and centroid motion as described above.

However, we found a better procedure with which we base our conclusions.
We apply the project's CBVs to our aperture photometry, using an altered goodness criterion for the coefficients to create a LC with normalized corrected counts (NCC): 

$$
NCC = \frac{Counts}{\langle Counts \rangle} - \Sigma_i C_i \times CBV_i
$$
   
A criterion based on minimum rms deviation from the mean can falsely accept a solution in which the corrected light curve oscillates around the mean more, yet fewer oscillations is preferred.
To understand this better, imagine a signal that rises near the beginning and then becomes constant and also imagine a single CBV that is just an upwards convex function, ie. does not match the shape at all.
Subtracting a multiple of the CBV could reduce the rms by a substantial amount by adding two more crossings of the average, but since it does not match the pattern of variability, it should be given a very small coefficient.   
What is required is a criterion that increases the coefficient if doing so improves the flatness of the corrected signal everywhere locally in time and not by having more zero crossings.  
We find good stable results with a criterion in which we minimized on the sum of the squares of the changes separated by any length from 10 - 25 days. 
If a shorter delay is used then the low frequency trends are not sufficiently diminished, and if a longer delay is used then the low frequency trends are over suppressed  at the expense of the high frequencies.  
We also limit the value of the coefficients that we apply to the normalized CBVs to $< 1.5$ to prevent over-correction on the normalized LCs. 
In the PDC procedure, these coefficients have been found to typically be under 1.0.
We start by using the first two CBVs and include an additional CBV if the decrease in the criterion was significant, up until a maximum of six CBVs.  The Quarterly CBVs provided by the project are ordered by the amount of correlation that they have with all of the LCs of targets on a given CCD of the camera.
   
After the correction are made to a Quarter, we sew pairs of Quarters together by fitting a few days at each Quarter's border with third degree polynomials.  
A time is selected several days into the later Quarter and the ratio of the two polynomials at that time is the constant multiplier applied to the later Quarter.

After about one year, Kepler returns to the same  Sun-Kepler-FOV configuration and the counts and the trending  pattern repeats fairly closely that of the previous year.
This can be useful for solving the long term trending puzzle.
We first use the $3 \times 3$ pixel apertures, and where we have observations over more than one year, we check that  the fluxes now repeat well even without sewing every 5th Quarter to its previous one.
If the yearly repeat is off by more than the error of a single exposure, then we redo the photometry with a $5 \times 5$ pixel aperture, and this usually works well.   
If there is still a problem, often it is an indication that we have overused the CBVs and, usually, reducing the number of CBVs utilized achieves good matches on the one year timescale.
%The bottom line is that a number of the galaxies that looked like they were active in the PSC light curves did not look active after they went through our reduction procedure.  
%Also, the resulting PSDs and SFs of active galaxies in our procedure are sometimes somewhat different from those from the PSC light curves.

\section{Newly Discovered Active Galaxies}
\begin{sidewaysfigure}
\includegraphics[scale=.6]{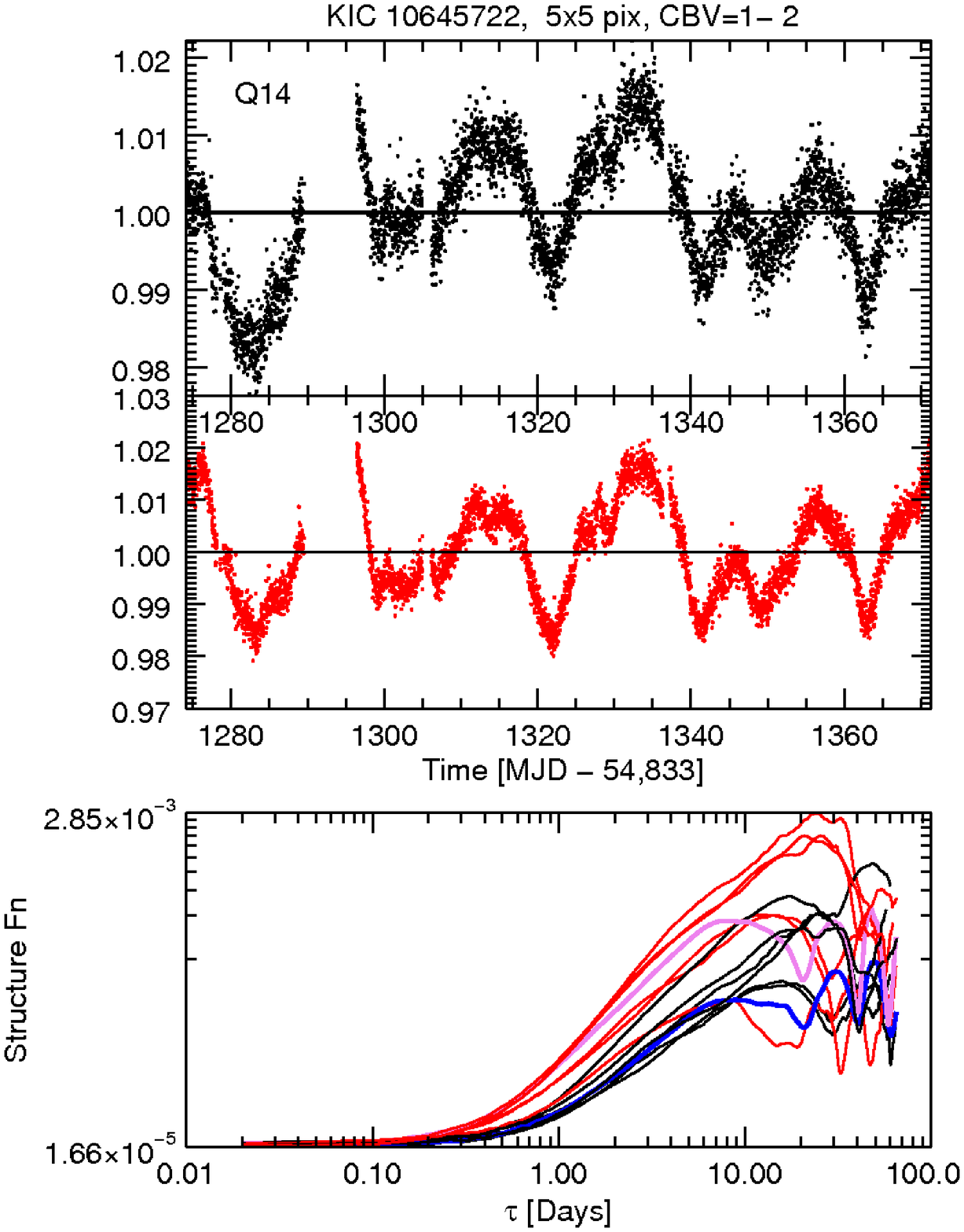}
\includegraphics[scale=.6]{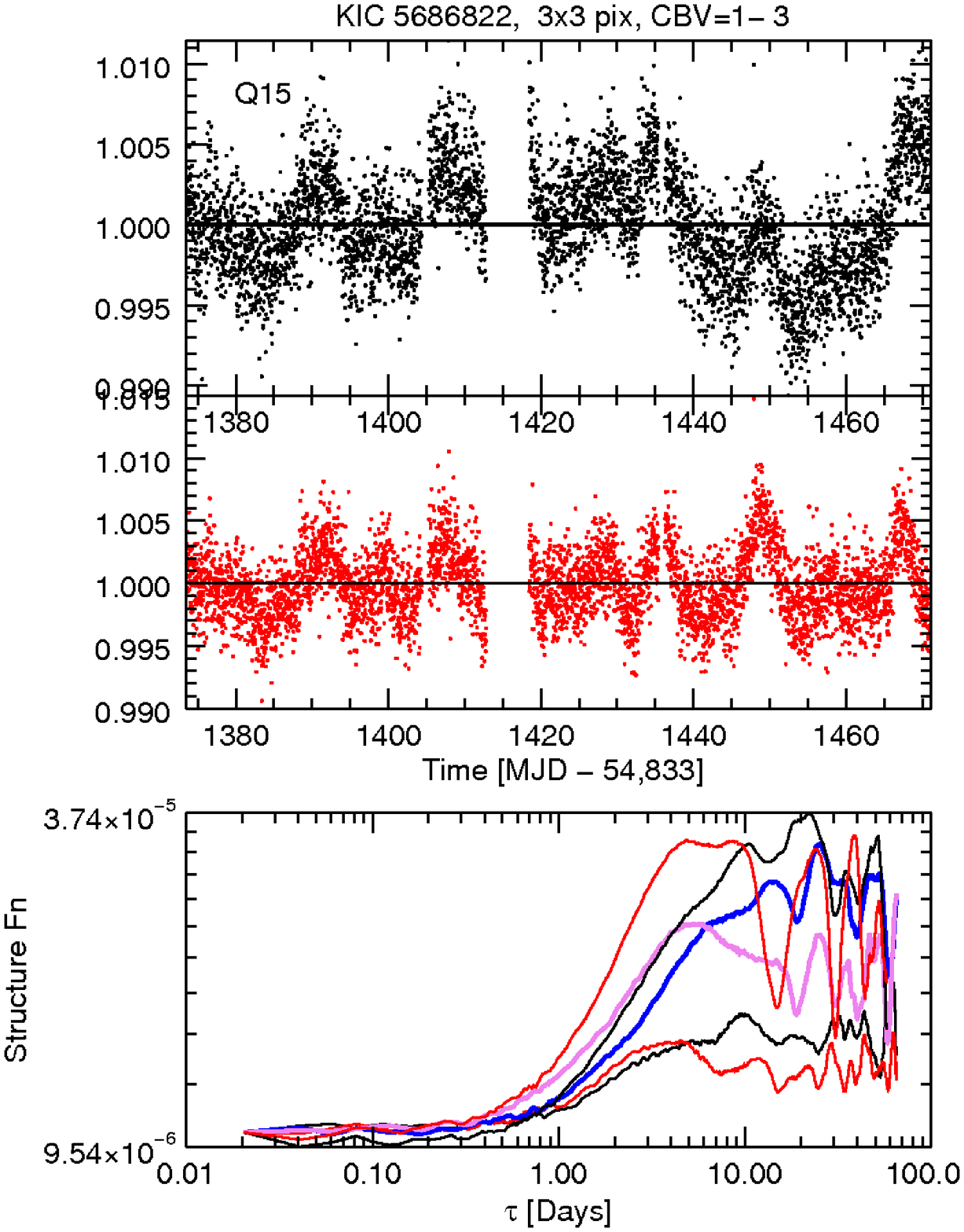}
\caption{Corrected and normalized single quarter light curves (black dots) of galaxies KIC 10645722 using a $5 \times 5$ pixel aperture (left) and the first two CBVs and KIC 5686822 (right) using a $3 \times 3$ pixel aperture and CBVs 1 - 3. The red dots are the project's PDC corrected light curves using 'optimal apertures'.  These two have strong enough activity to be obvious AGN candidates.  The former needed just two CBVs for long term detrending because the amplitude of its variation was high.    Below the light curves are structure functions from the various quarters where the target was observed and the detector was well behaved.  For the quarter shown above,  we use blue lines for our analysis and violet  for the PDC, but for other quarters we use black and red.  \label{active1}}
\end{sidewaysfigure}

After correcting for long term trends and eliminating Quarters where CCD bias artifacts were dominant, only 4 active galaxies were easily discernible by visual inspection of the all the light curves (seen in Fig.~\ref{active1} and \ref{active2}).  
To detect lower amplitude active galaxies, we made use of the fact that for known
AGN,  typically, the PSDs, fit on the timescale of weeks or longer, have power laws going as $f^{-\alpha}$ with $\alpha$ between 1.5 and 2.5.  Thus, the ratio of the power at many days to that at just a few hours could be a discriminator for detecting AGN.

\begin{figure}
\epsscale{1.0}
\plotone{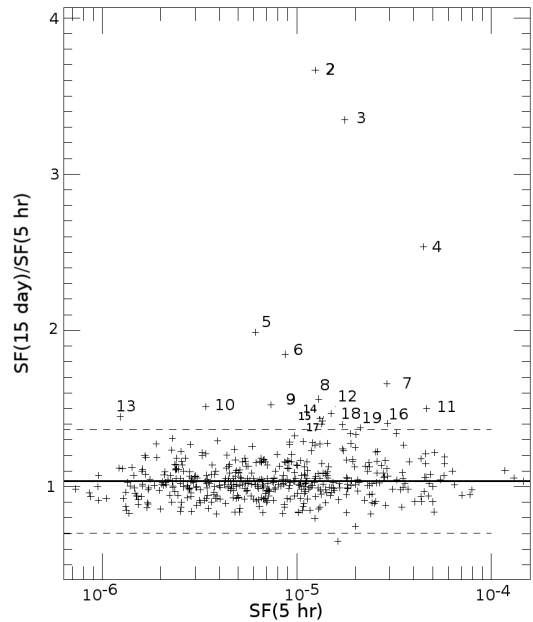}
\caption{Ratio of Structure Function at two delay times, 15 days and 5 hours, as a function of Structure Function at 5 hours for 457 galaxies measured by Kepler Mission.  The candidate AGN are numbered 1 to 19 (identified in Table 1) and have ratios $> 3 \times$ the standard deviation of the distribution (dashed lines), $\sigma = 0.11$. The most active galaxy lies off the plot at a ratio of 29.9.
 \label{sf5h15d}}
\end{figure}
To obtain a statistically valid estimate of the active galaxy occurrence rate, we devised an automated procedure that would resist being triggered by temporary periods of high instrumental noise occurring in many of the CCDs.  
The sample was restricted to the subset of targets with three or more Quarters of observations; 473 galaxies remained in the set.  
For each of these, structure functions were made on the quietest Quarter of the PDC LC; this ensures that none are being selected as active candidates because of excess instrumental noise.  
Although this precaution may result in excluding active galaxies that happened to be very quiet during one 3-month period, we believe that this is rare, while the danger of being fooled by instrumental noise was high. 
We defined a dimensionless, unweighted structure function, 
$$
\mathcal{SF}( \tau )= \frac{\langle [F(t+ \tau ) - F(t)]^2\rangle}{\langle F(t) \rangle ^2}.
$$
where $F(t)$ is the brightness at time $t$, $\tau$ is a delay time, and angle brackets mean average over all points in the time series.
We first smooth the data by averaging sets of four 30 minute exposures, in this form, the structure function directly provides the mean square of the fractional change at different delay times in 2 hour flux measurements.
The angle brackets indicate an average is taken over the differences in all pairs in the LC at a given delay time.

\begin{deluxetable*}{rrrrrrrrc}
\tablecolumns{9}
\tablewidth{0pc}
\tablecaption{Active Galaxies Discovered in sample of 474 galaxies}
\tablehead{
\colhead{Rank}&
\colhead{Kepler ID} &
\colhead{RA}&
\colhead{Dec}&
\colhead{g}&
\colhead{r}&
\colhead{J}&
\colhead{Skygroup}&
\colhead{SF(10d)/SF(4h)}}
\startdata
       1 &   10645722&18 47 22.340&+47 56 16.13&  15.958& 15.635 &15.505&11&29.95  \\ 
       2 &     5686822&18 59 12.137&+40 59 26.73&  16.801& 15.872&15.520&50& 3.66  \\ 
       3 &  11716536&19 35 17.046& +49 51 51.67&  17.063& 16.218&16.302 &14& 3.35  \\ 
       4 &    7986325&19 57 55.472& +43 46 49.27&  16.850&\nodata&15.456 &53& 2.54  \\ 
       5 &    6751969&18 47 30.765&+42 17 07.23&  16.002&  15.515&15.594&31& 1.99  \\ 
       6 &    7691427&19 38 47.414&+43 21 57.31&\nodata&  17.114&15.080&58& 1.85  \\ 
       7 &  12553112&19 12 52.522&+51 35 06.37&  17.844&  16.666&15.831& 3& 1.66  \\ 
       8 &  11768473&19 38 04.715&+49 55 47.89&  16.222&  15.722&15.936 &14& 1.56  \\ 
       9 &    2142191&19 06 03.085&+37 35 09.70&  15.696&  14.759&15.021&71& 1.52 \\ 
      10&   12556836&19 22 00.693&+51 35 56.09&  14.785 &  14.321&14.939& 4& 1.51  \\ 
%NVSS J192200+513557 (3.7 mJy)\\
      11&  11808151&19 14 16.560&+50 00 24.84&  18.229&  17.251&16.356& 2& 1.50  \\ 
      12&    5683305&18 52 03.013&+40 59 33.73&  17.005&  16.274&15.714&52& 1.47  \\ 
      13&   8024526&19 21 25.589&+43 53 16.90&  14.971&  14.393&14.282&41& 1.45  \\ 
% z=0.0551
      14&   5511084&18 51 55.133&+40 47 23.88&  14.984&  13.925&14.442&52& 1.44  \\ 
      15&   6714622&19 50 09.928&+42 06 53.09&  17.544&  16.228&15.064&74& 1.42  \\ 
      16&   4148802&19 16 58.775&+39 14 48.41&  18.802&  17.757&15.789&68& 1.40  \\ 
      17&  10402746&19 24 07.041&+47 33 41.50&  16.991&  16.245&15.855&18& 1.40  \\ 
      18&    9509125&18 50 29.335&+46 08 01.14&  18.376&  17.357&15.487&10& 1.40  \\ 
      19&    8884097&19 25 08.969&+45 10 29.57&\nodata&  17.699&15.601&43& 1.38  \\ 
\enddata

 %      6 & 09875289 & 18 54 22.925 & +46 43 15.74 & 18.424 & 17.504 & 15.672 & 29 &  1.82 \\
%  8 & 10199926 & 19 13 47.961 & +47 15 21.93 & 15.772 & 15.260 	& 15.939 & 21 &  1.69 \\
%       9 & 05683305 & 18 52 03.013 & +40 59 33.73 & 17.005 & 16.274 & 15.714 	&   4 &  1.69 \\

\end{deluxetable*}

For galaxies with no activity, the SF should be a constant at all delay times. 
As an automated procedure to discriminate active from quiet galaxies, we form the ratio of the SF with delay times at 5 hours to that at 15 days, averaging every four time samples to reduce noise.  
This short delay time is almost as short as we can go with the averaging used.  
At these particular two points in the delay times, the number of standard deviations from the mean in this ratio are maximized for these candidates and subsequently the maximum number of candidates is found.
Figure~\ref{sf5h15d} shows the distribution of this ratio as a function of the SF at 5 hours.  
The mean of the ratio was 1.03 with root mean square deviation of $\sigma$ = 0.11.  
Of 474 galaxies, 19 galaxies have ratios $>3\sigma$, where $\sigma$ is the root mean square deviations of the ratios with $5\sigma$ deviants removed,  and these are labeled in the diagram with their rank order.  
In a normal distribution, only 0.15\% of the sample would be greater than $3 \sigma$ from the mean, which for this size sample corresponds to 0.7 galaxies.

Presently we do not know much about these AGN candidates.  
Searches on NED and HEASARC show that all but two of the 19 candidate AGN are unstudied other than by 2MASS measurements.   
None have detected x-ray emission.  KIC 12556836, also known as NVSS J192200+513557, was observed to emit $4.2 \pm 0.5$ mJy at 1.4 GHz in the NRAO VLA Sky Survey \citep{Condon_etal98}, but no spectra have been taken. 
KIC 8024526 was determined to have a redshift of z = 0.0551 (m - M = 36.80, H$_0$ = 72 km/s/Mpc)   \citep{FaberDressler77} and thereby confirmed to be a member of Abell Cluster 2319.  
 The absolute magnitude for the whole galaxy is M$_r$ = -22.4 mag ($3.8 \times 10^{10} L_\odot = 2.2 \times 10^{44}\ ergs/s$), and so the detection of variability at the level of 0.5 milli-magnitudes corresponds to luminosity variations of $1.1 \times 10^{41}$ ergs/s at 230 Mpc. 

KIC 10645722 (the most active candidate, Fig.~\ref{active1}) was selected to be studied spectroscopically at Lick on the basis of its variability on Kepler and having infrared colors from WISE in the region predominantly populated by AGN \citep{EdelsonMalkan2012}.  
There is an r-band CFHT Megacam image of it, and it was imaged in Cycle 21 on HST with the WFC/UVIS (HST proposal 13409) .  It is a Seyfert 1 at redshift z=0.068 (K.L. Smith, personal communication) or distance modulus 37.26 (Ho = 72 km/s/Mpc).  
This galaxy has a total luminosity of $1.5 \times 10^{44}$ ergs/s and HST and MegaCam images show that the majority of its light comes from the central source.
 
Variations are seen,  over three years, having amplitude of $\pm$ 4\% for this source, while other candidates vary by  $< \pm$ 0.5\%.
Assuming the BH is emitting below the Eddington Limit ($L_{Edd} = 1.3 \times 10^{38} M_{BH}/M_\odot$ erg/s), the minimum mass of the BH is $\sim1.1 \times 10^6 M_\odot$.

 If we identify elliptical galaxies as those having photometric color $r-J > 1.0$ mag, then four of the 19 are ellipticals:  KIC~7691427, 6714622, 4148802, 9509125, and 8884097.  Interestingly, these are all toward the lower end of the ranking in the detection ratio with 3 of the 5 in the bottom 4.   Given the favorable weighting of elliptical galaxies in the selection criterion for our targets, this indicates a strong preference for detecting spiral galaxies as active. 

Unfortunately, we do not have redshifts yet for most of these galaxies, so we cannot convert their brightnesses to luminosities.

\subsection{Structure Functions and PSDs}

To characterize the variations as a function of timescale, we use the unweighted Structure Function defined above.  
SFs are stable if the time series contains many gaps or long gaps, as long as the data are not too sparse.  Small gaps exist in this data from several causes, chiefly: cosmic ray events, several day events to turn the spacecraft and download the data to earth on a monthly basis, Quarterly rotations of the spacecraft, Moire pattern noise and other electric interference events on the CCD.  Sometimes a given target was simply not scheduled for a downlink for a specific Quarter.  

%The SFs can be fit with power laws just as power-spectra density functions are.  In a system %with a pure power law PSD of index $\alpha$, the Structure Function would have an index %$\beta$ which is related to $\alpha$ by $\alpha = 1 + 2\beta$ \citep{Kawaguchi_etal98}.

\begin{sidewaysfigure}
\includegraphics[scale=.6]{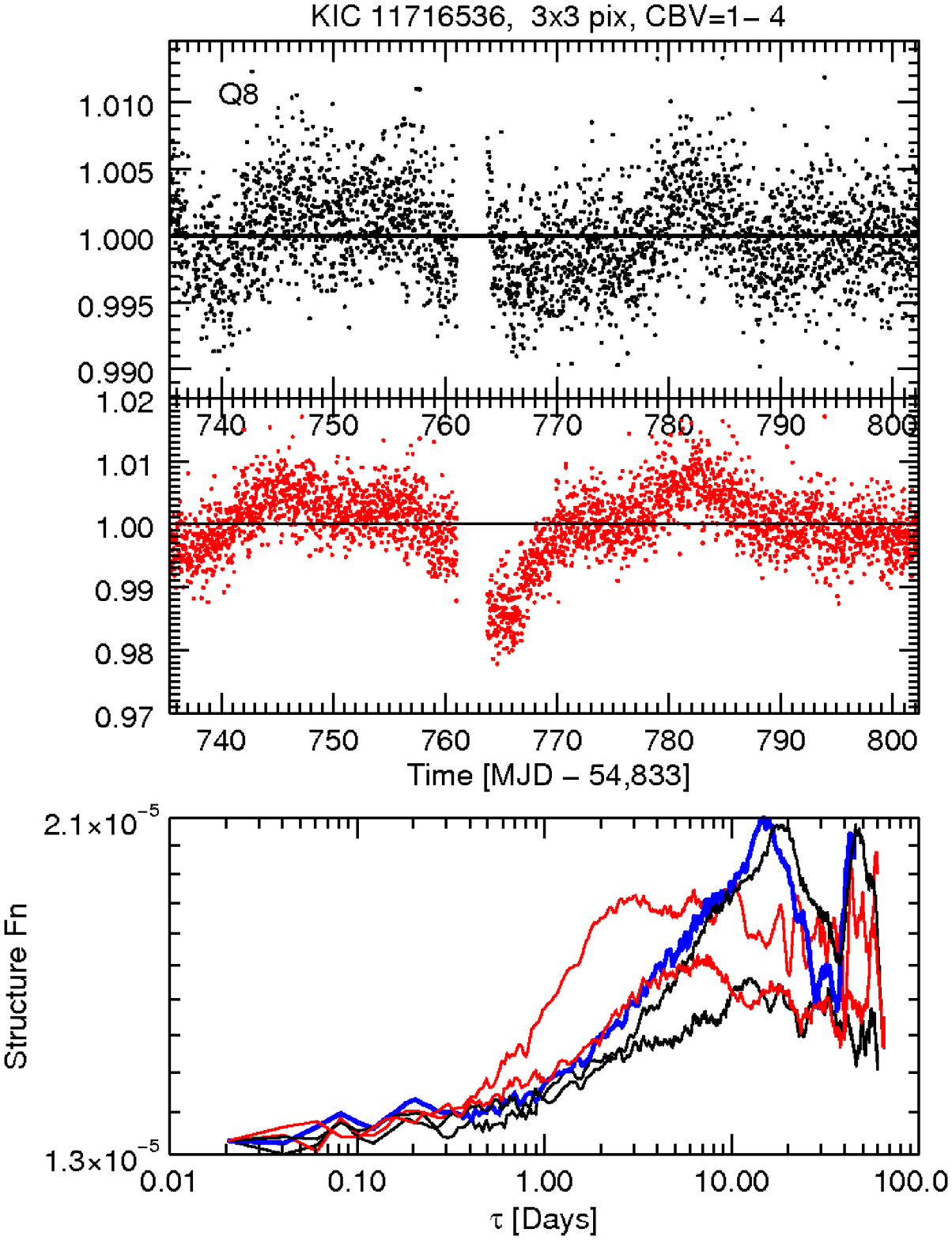}
\includegraphics[scale=.6]{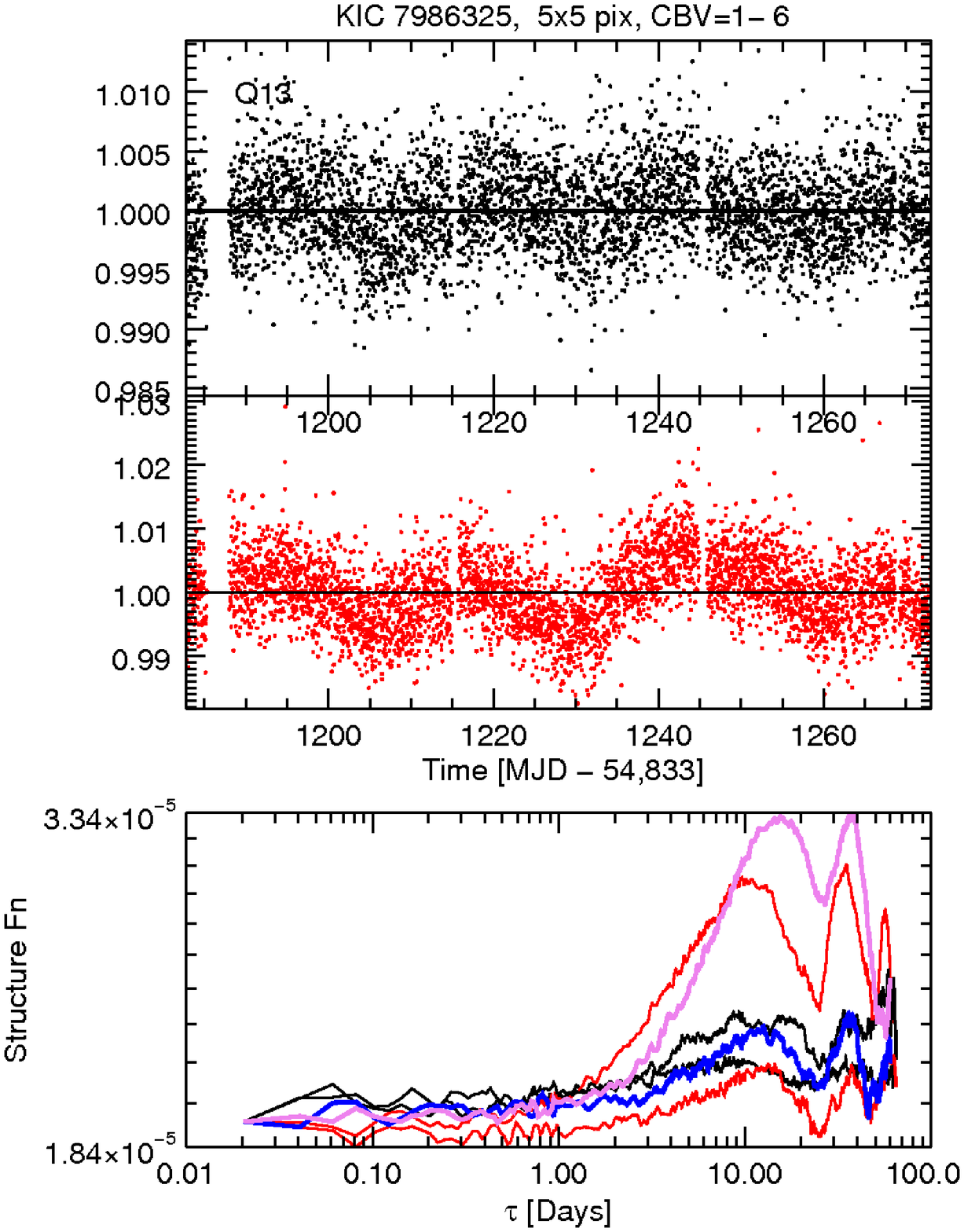}
\caption{Corrected light curves (top) and Structure Functions (bottom) from our analysis (blue and black) and the project's PDC (red). KIC 11716536 (left) and KIC 7986325 (right). \label{active2}}
\end{sidewaysfigure}

\begin{sidewaysfigure}
\includegraphics[scale=.6]{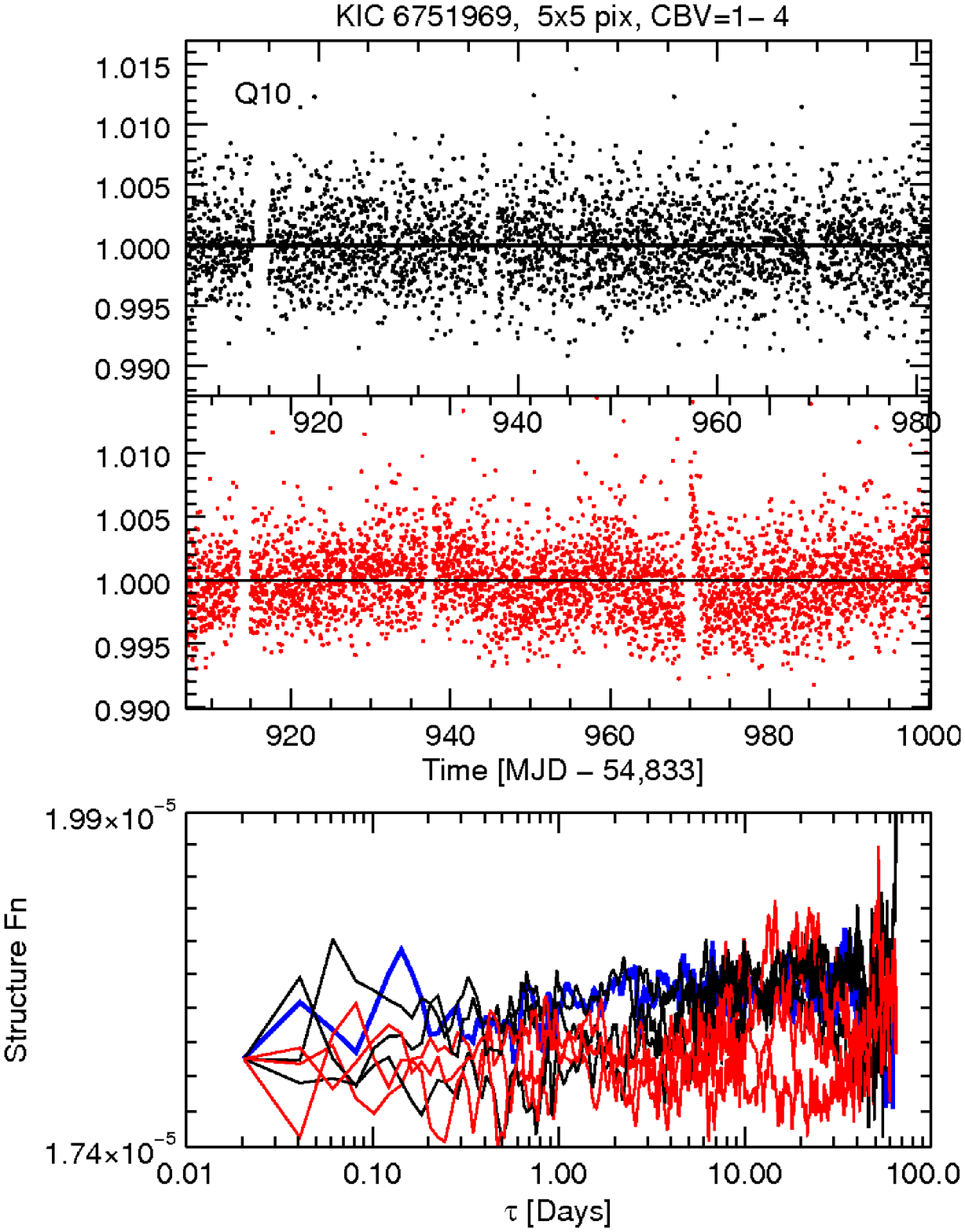}
\includegraphics[scale=.6]{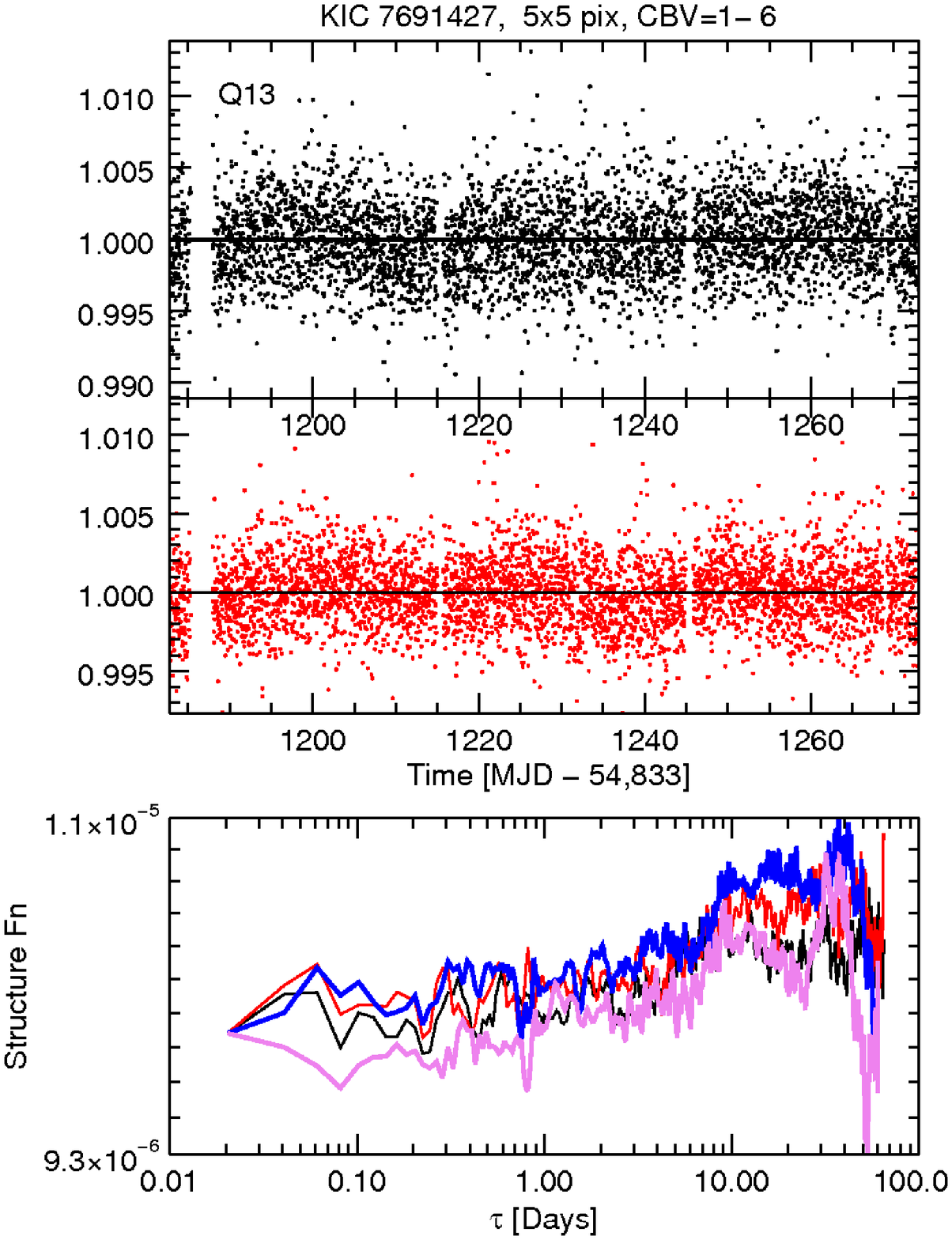}
\caption{Corrected light curves (top) and Structure Functions (bottom) from our analysis (blue and black) and the project's PDC (red). KIC 6751969 (left) and KIC 7691427 (right). \label{active3}}
\end{sidewaysfigure}

\begin{sidewaysfigure}
\includegraphics[scale=.6]{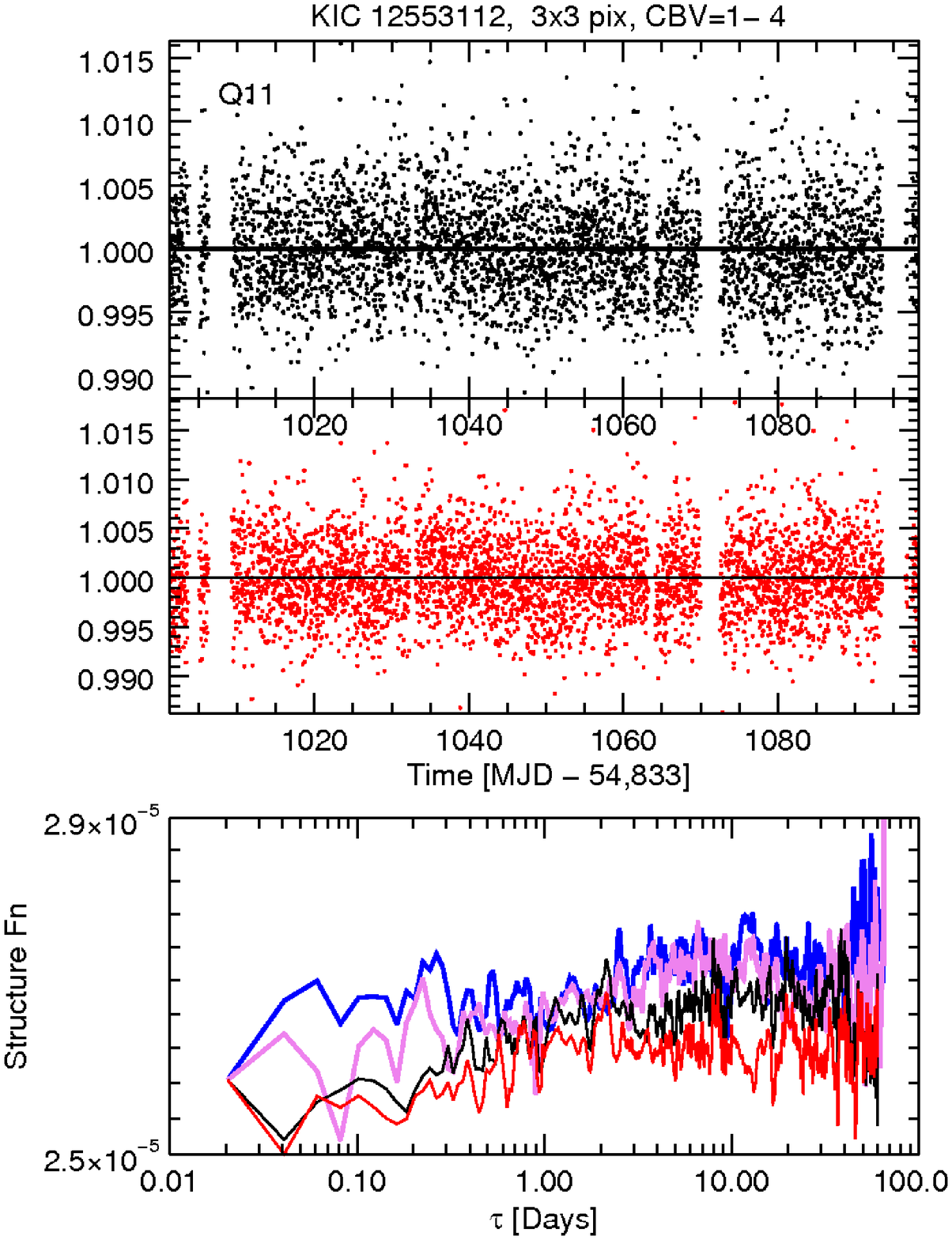}
\includegraphics[scale=.6]{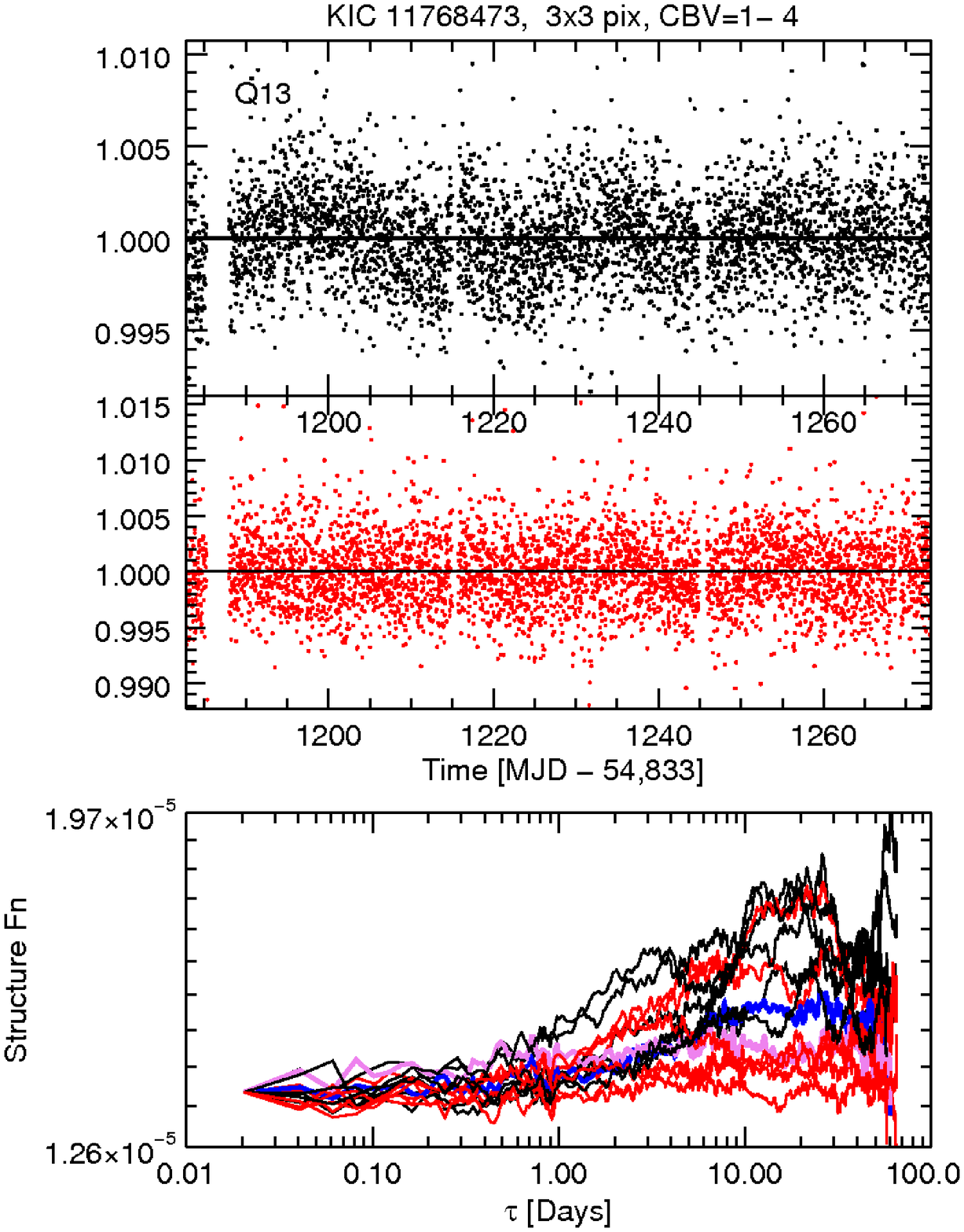}
\caption{Corrected light curves (top) and Structure Functions (bottom) from our analysis (blue and black) and the project's PDC (red). KIC 12553112  (left) and KIC 11768473 (right). \label{active4}}
\end{sidewaysfigure}

\begin{sidewaysfigure}
\includegraphics[scale=.6]{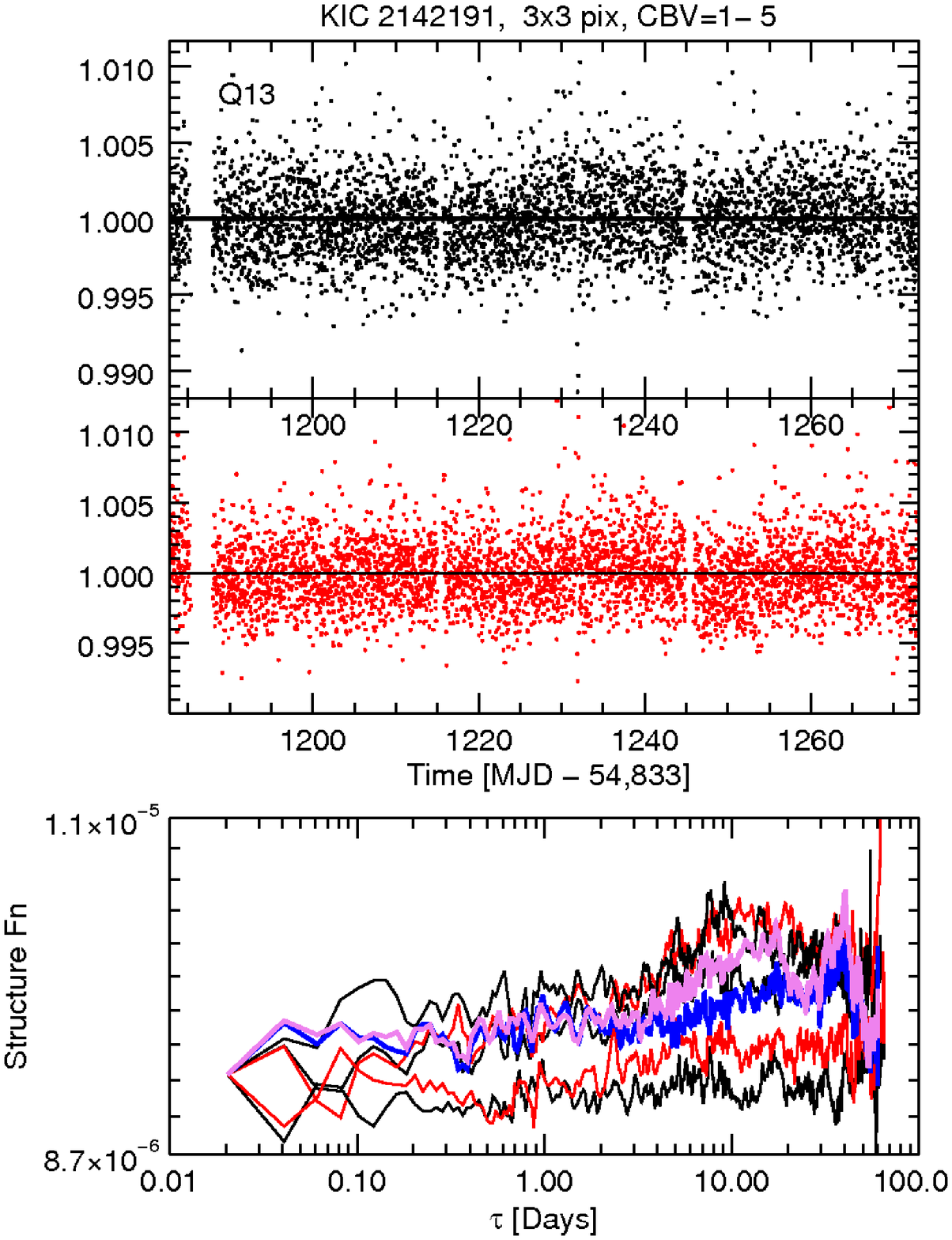}
\includegraphics[scale=.6]{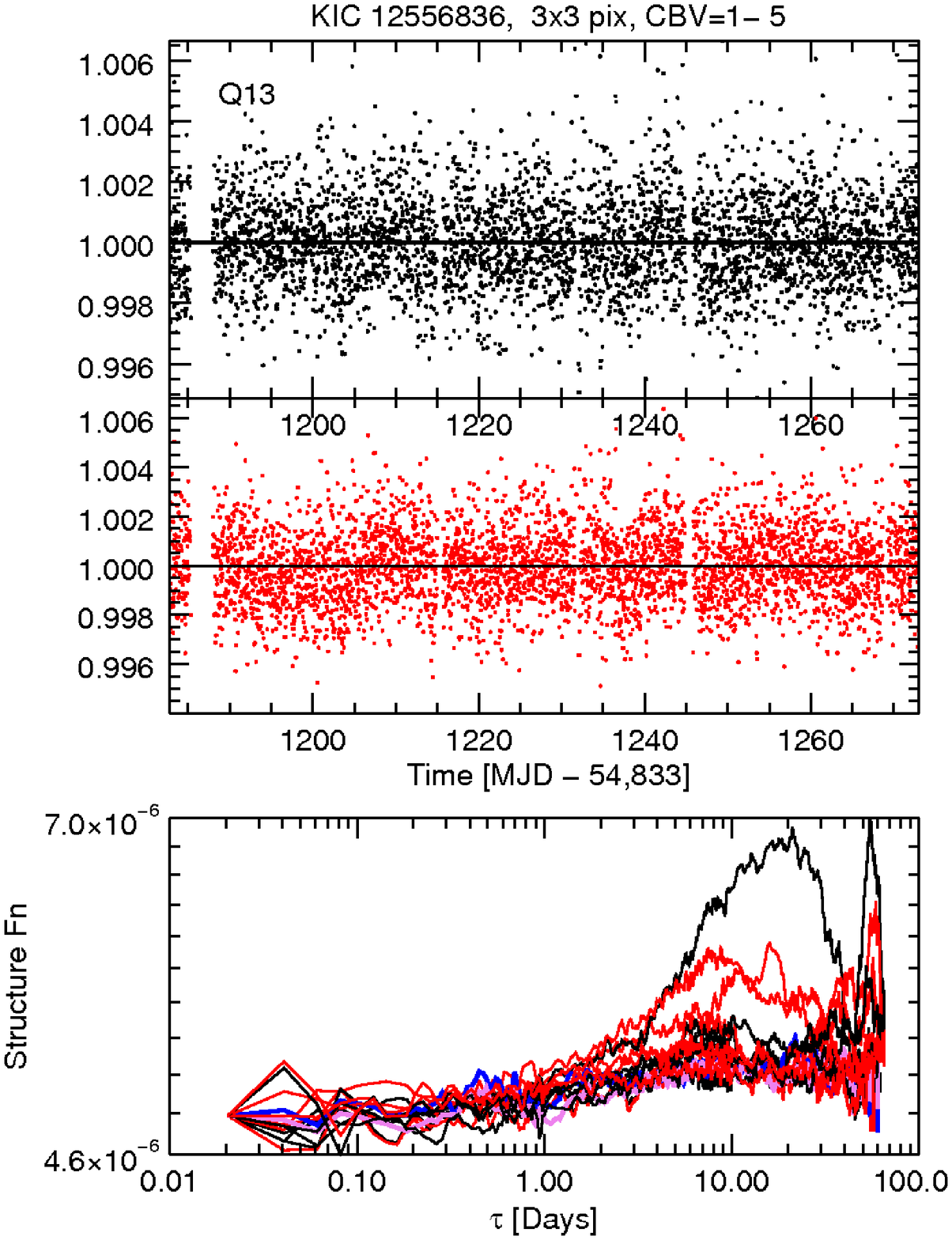}
\caption{Corrected light curves (top) and Structure Functions (bottom) from our analysis (blue and black) and project's PDC (red).  KIC 2142191 (left) and KIC 12556836 (right).  \label{active5}}
\end{sidewaysfigure}

\begin{sidewaysfigure}
\includegraphics[scale=.6]{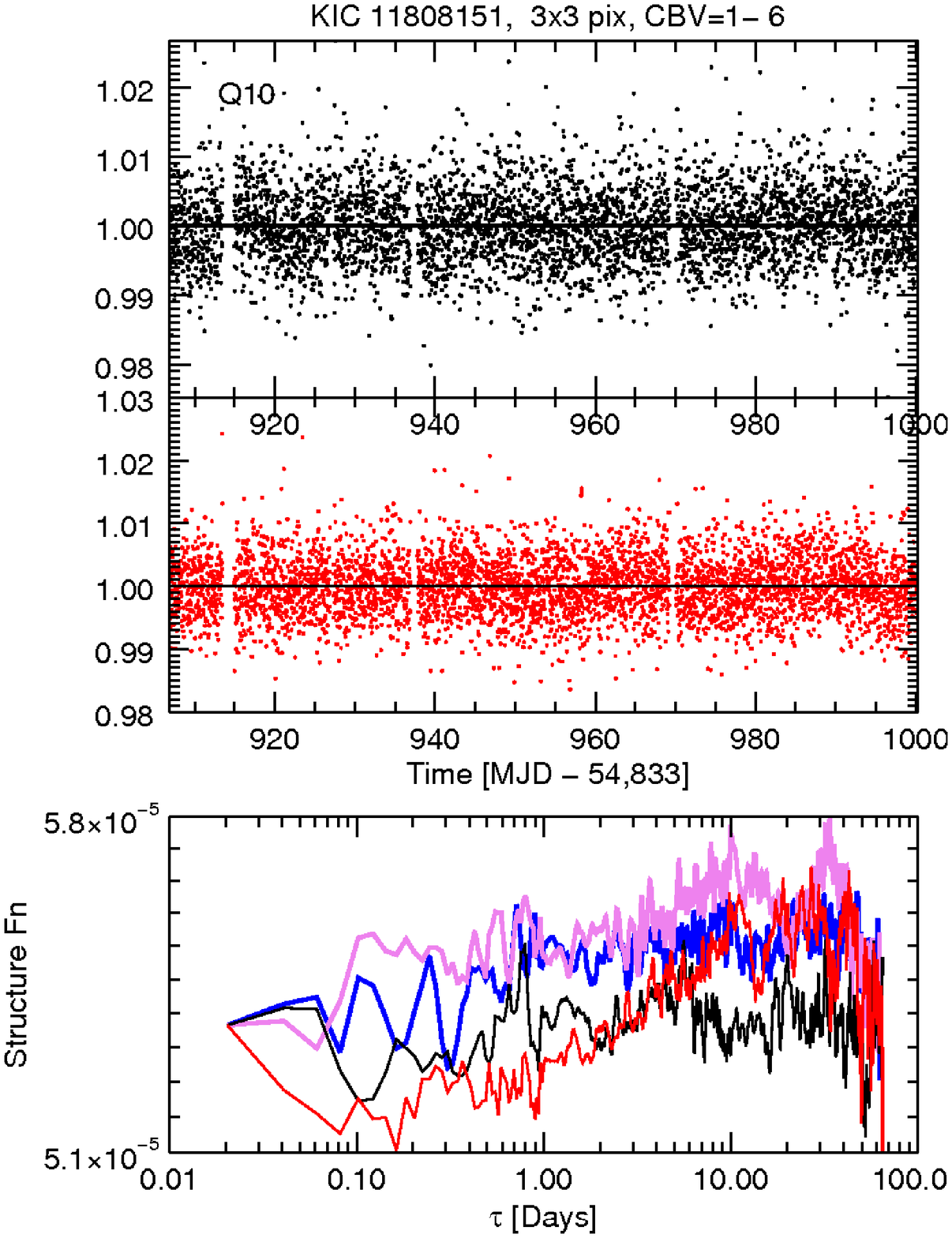}
\includegraphics[scale=.6]{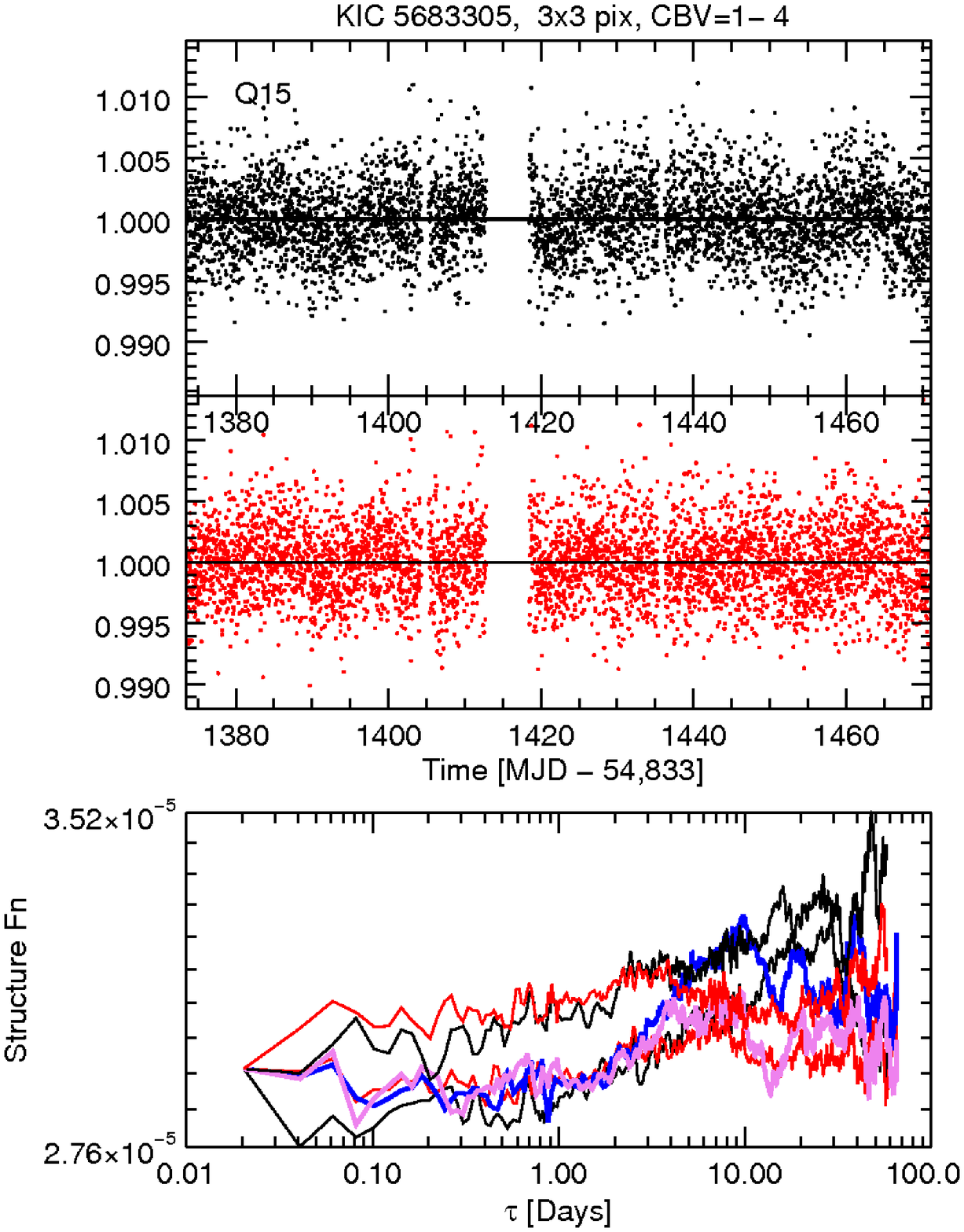}
\caption{Corrected light curves (top) and Structure Functions (bottom) from our analysis (blue and black) and project's PDC (red). KIC 11808151 (left) and KIC 5683305  (right).  Our reduced light curve of KIC 11808151 has less growth in fluctuations at longer delays the the PDC curve.  Therefore, this is not a likely candidate. \label{active6}}
\end{sidewaysfigure}

\begin{sidewaysfigure}
\includegraphics[scale=.6]{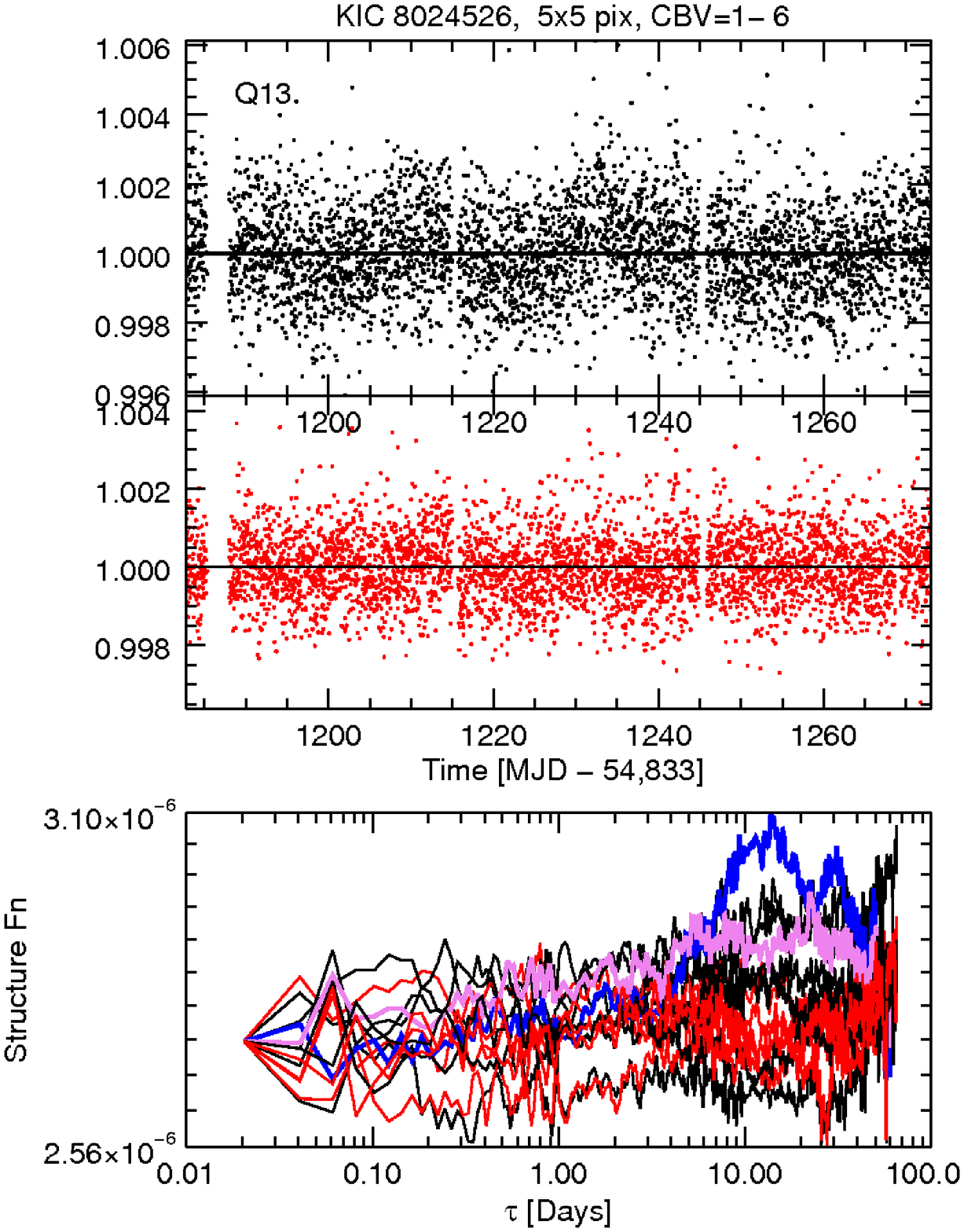}
\includegraphics[scale=.6]{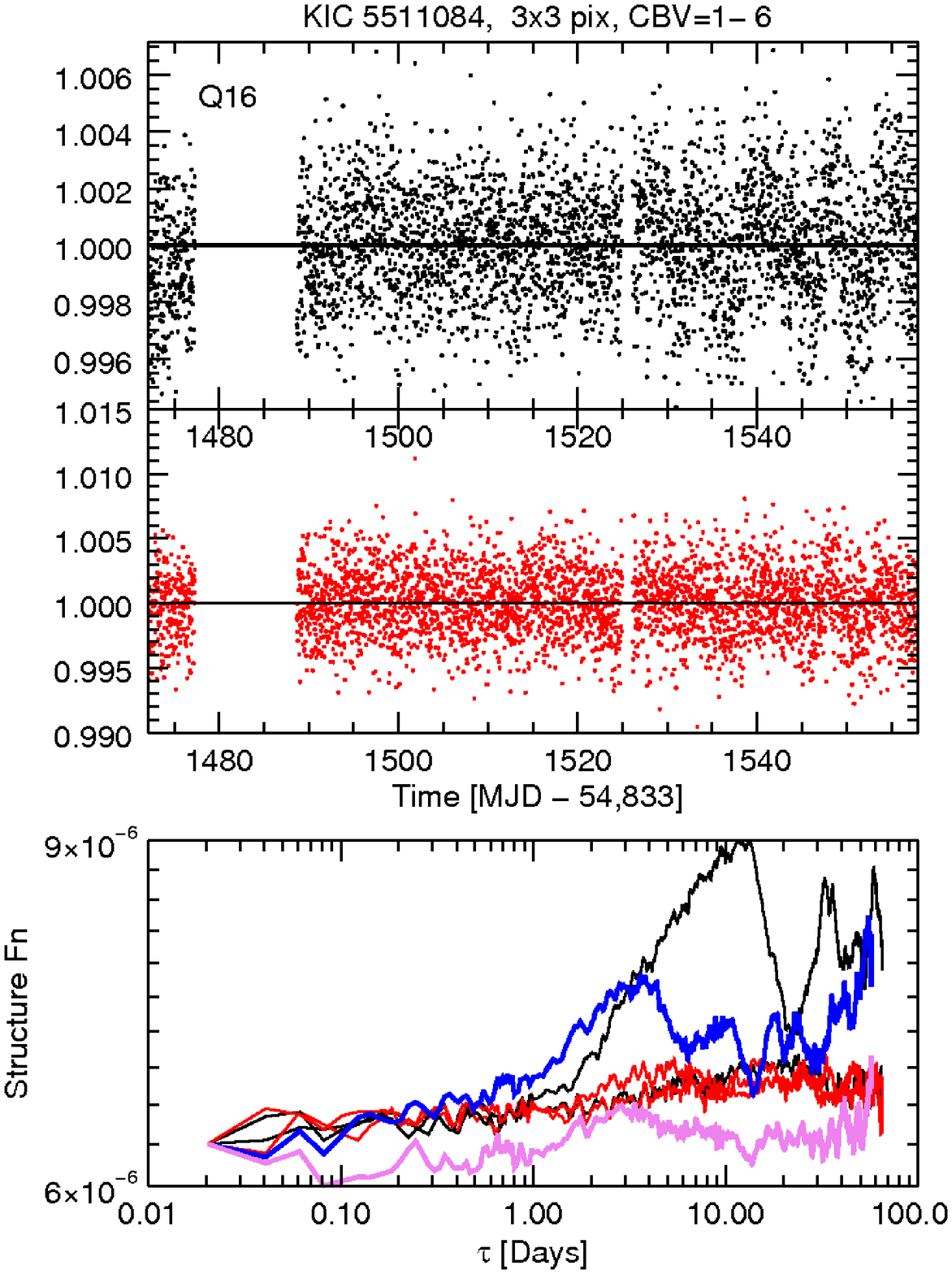}
\caption{Corrected light curves (top) and Structure Functions (bottom) for our analysis (blue and black) and project's PDC (red). KIC 8024526 (left) and KIC 5511084  (right). KIC8024526 had a glitch in Q9, the quietest rms Quarter.  However,  Quarter 11 is quiet, which indicates that this galaxy is not active.  \label{active7}}
\end{sidewaysfigure}

\begin{sidewaysfigure}
\includegraphics[scale=.6]{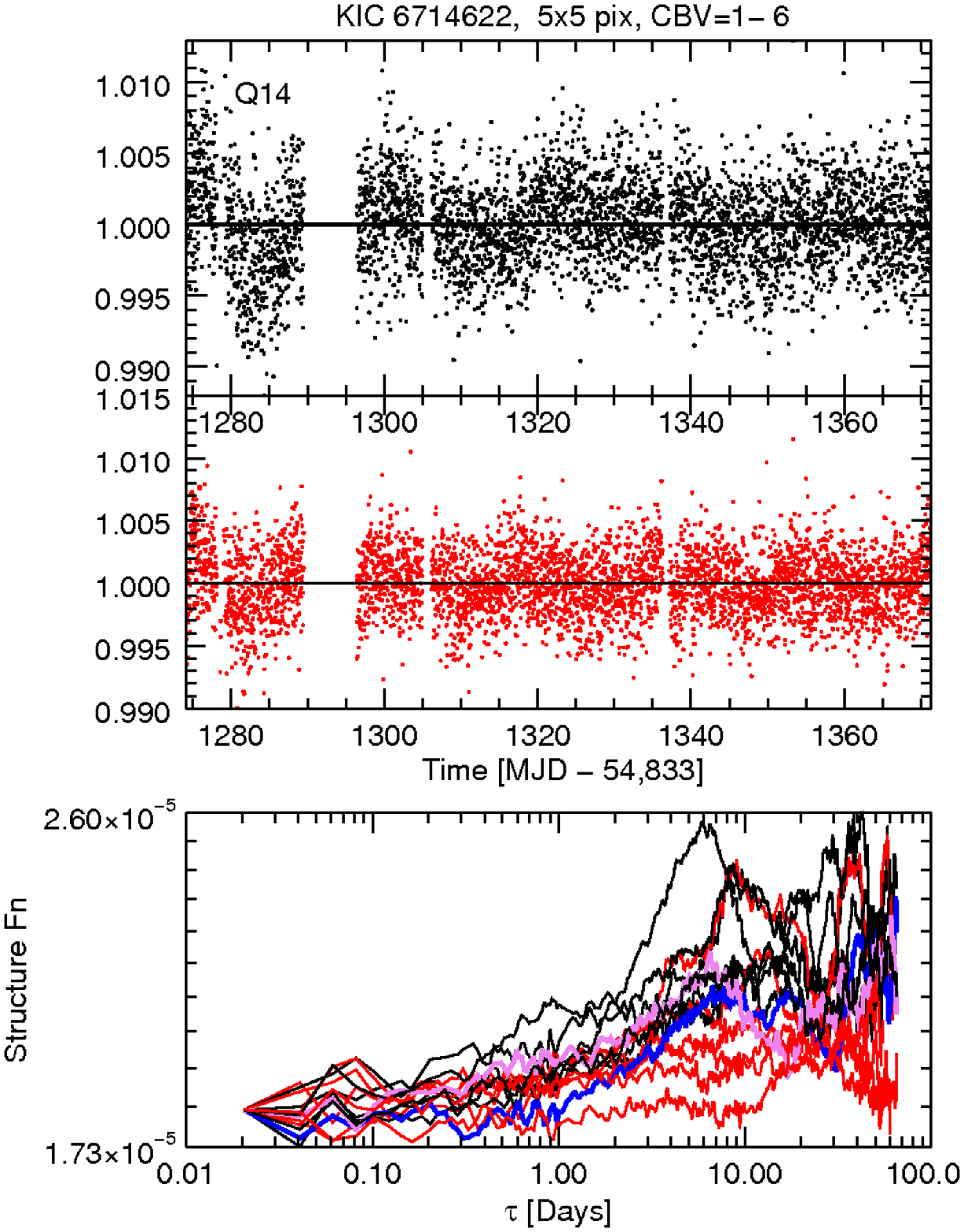}
\includegraphics[scale=.6]{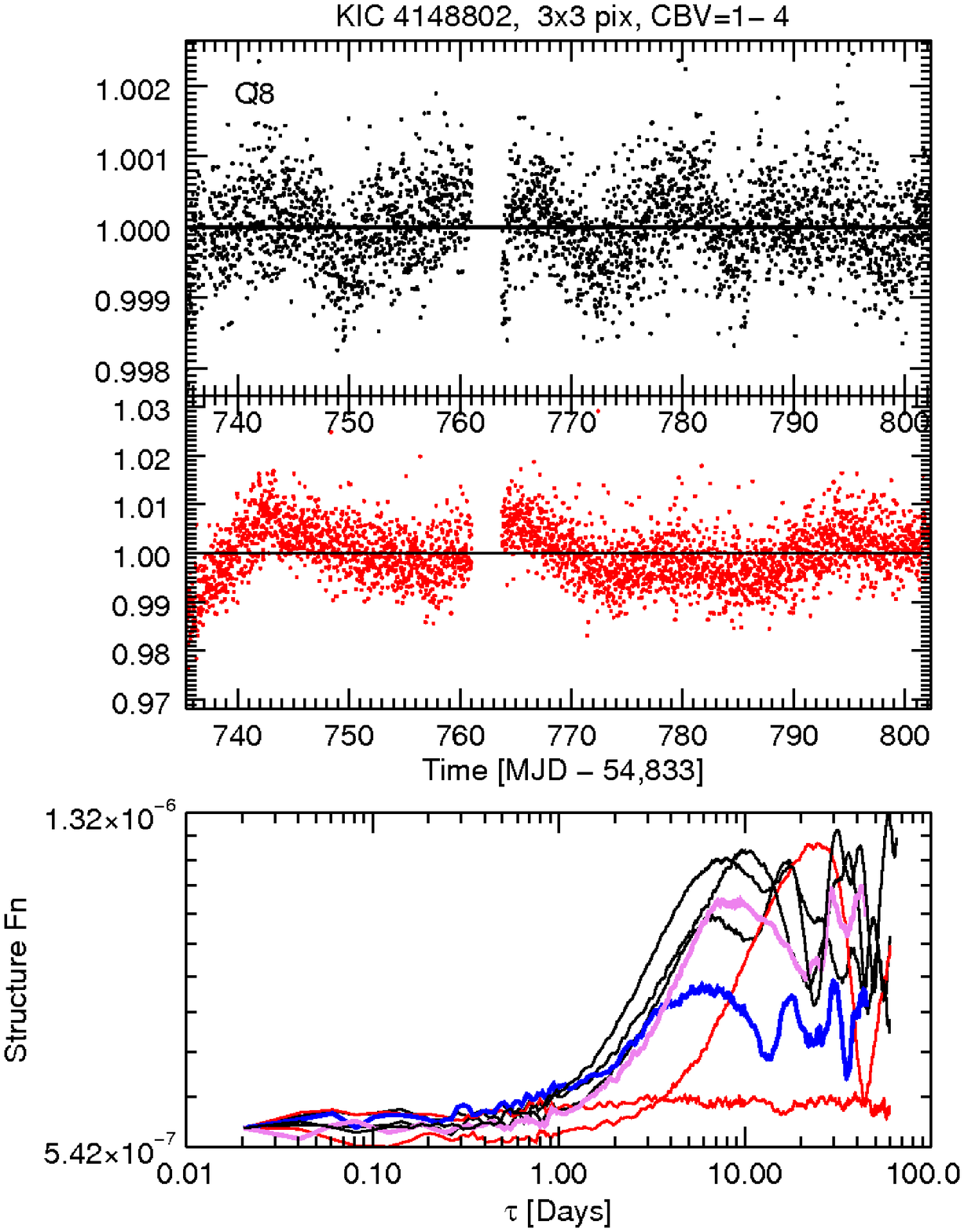}
\caption{Corrected light curves (top) and Structure Functions (bottom) for our analysis (blue and black) and project's PDC (red).  KIC 6714622  (left) and KIC 4148802 (right)The variability in light curves of both KIC4148802 and KIC6714622 were too suppressed by the pipeline processing.  For KIC6714622 we were able to make use of one year repeatability. \label{active8}}
\end{sidewaysfigure}

\begin{sidewaysfigure}
\includegraphics[scale=.6]{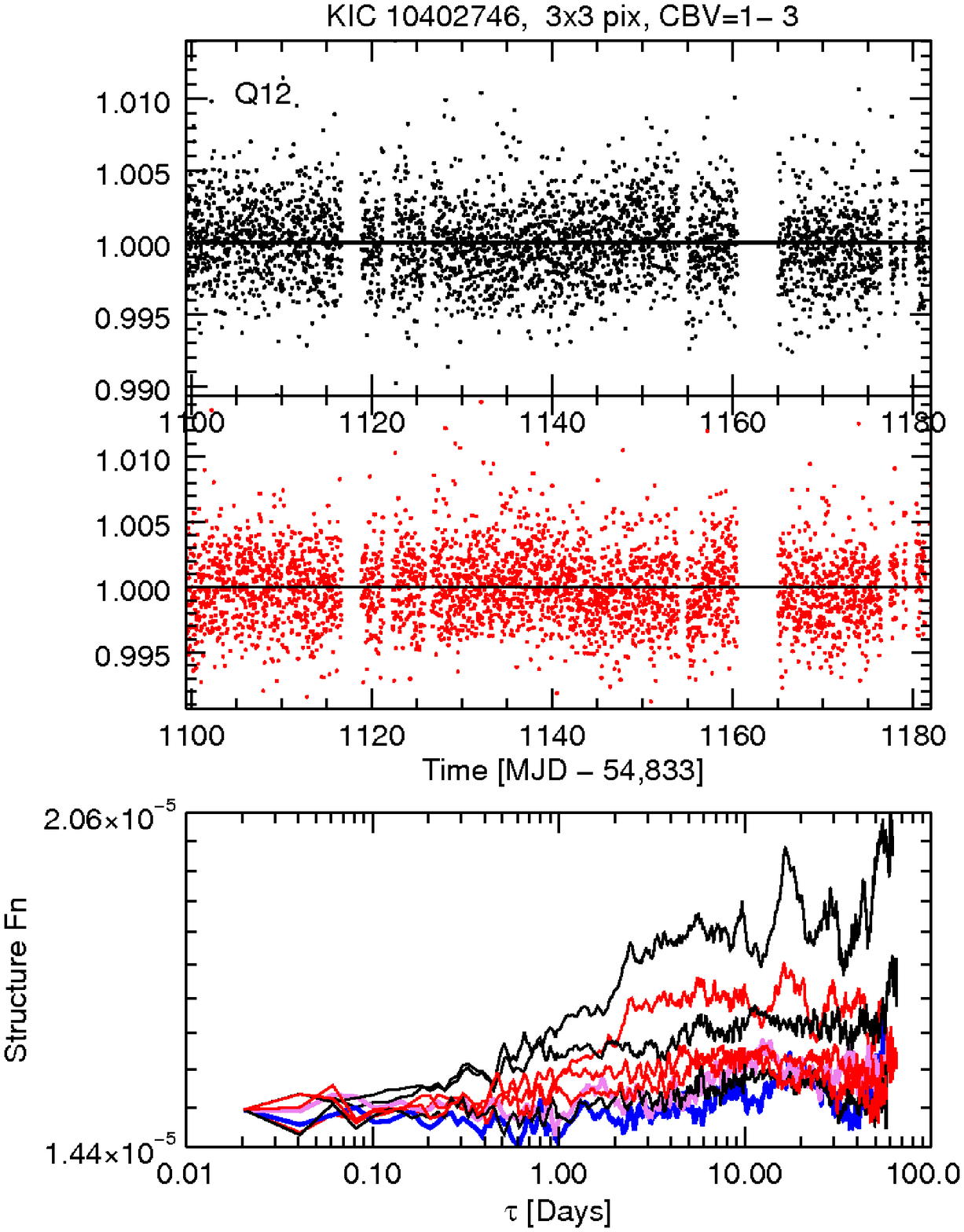}
\includegraphics[scale=.6]{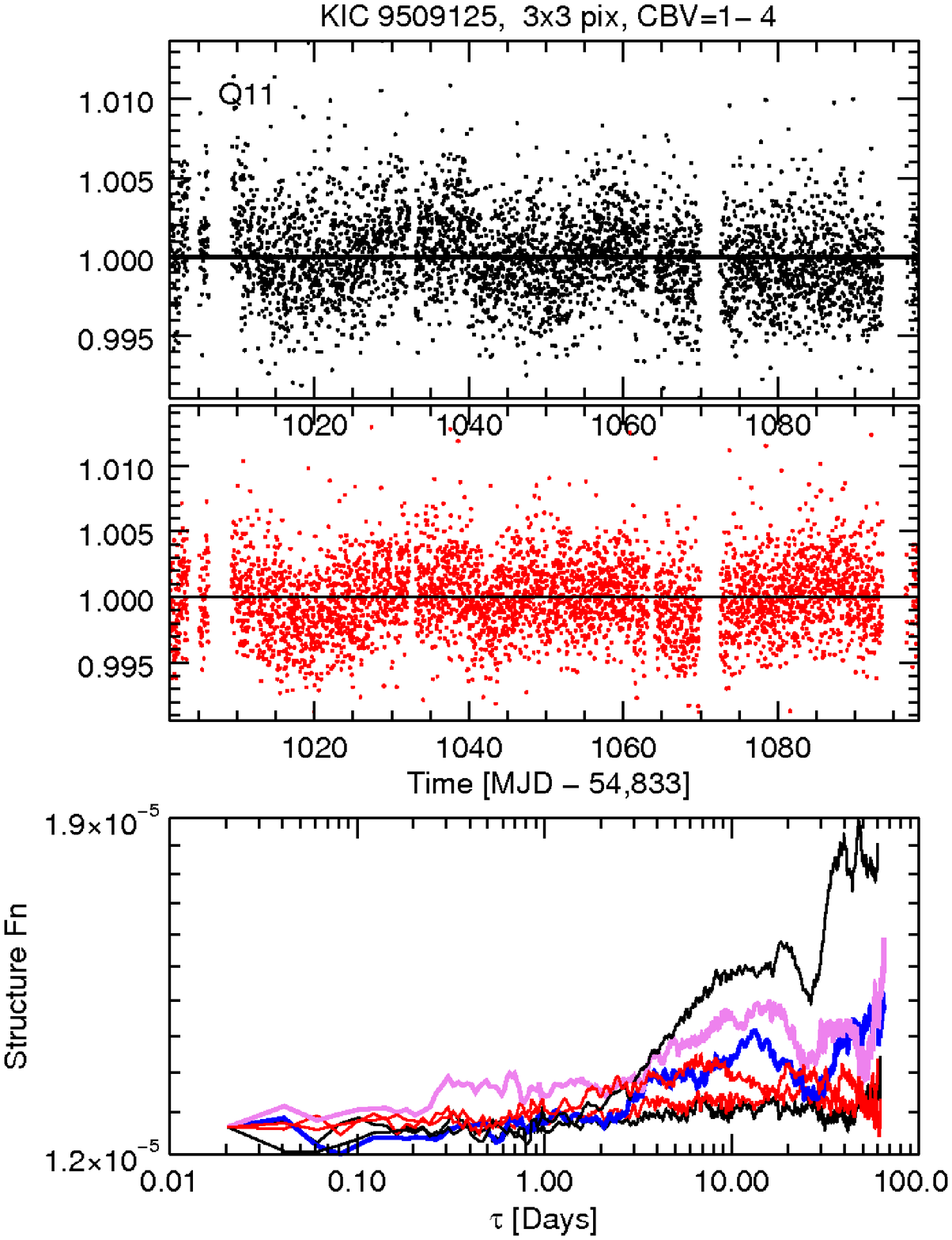}
\caption{Corrected light curves (top) and Structure Functions (bottom) for our analysis (blue and black) and project's PDC (red).  KIC 10402746 (left) and KIC 9509125  (right)  KIC10402746 is inactive in Quarter 16. \label{active9}}
\end{sidewaysfigure}

\begin{sidewaysfigure}
\includegraphics[scale=.6]{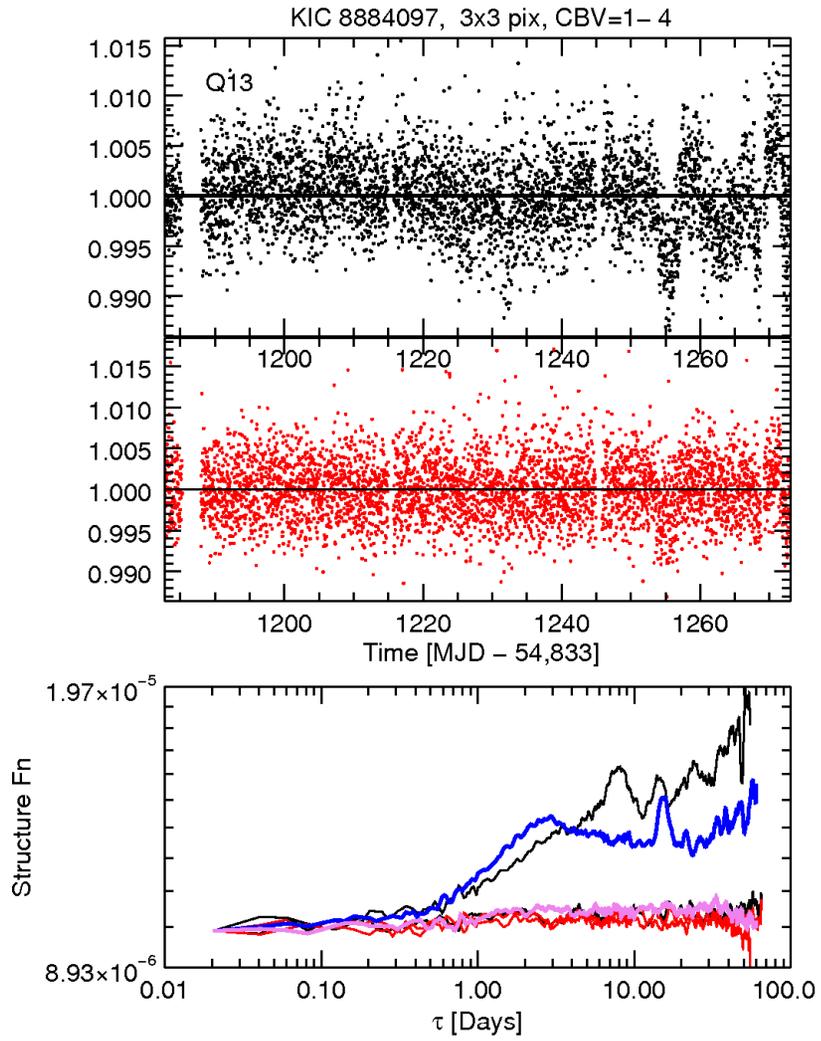}
\caption{Corrected light curves (top) and Structure Functions (bottom) for our analysis (blue and black) and project's PDC (red) for KIC 8884097. \label{active10}}
\end{sidewaysfigure}

The top two panels of  Fig.~\ref{active1} - \ref{active10} show LCs of representative Quarters for each candidate. 
The LCs,  after our processing  (black lines) differ by varying degrees from the Kepler project's processed LCs  (red lines).   
In some cases, the PDC failed completely or nearly so in the face of large variability.   
The SFs for all quarters with acceptable quality data are plotted (black for our reduction, red for the project's PDC, and blue and violet for the Quarter of the LC presented) in the lowest panel of these figures.  
 However, if the SF of the PDC LC is unreasonably noisy compared to our LC, it is not plotted.
The SFs are normalized to have the same amplitude at the highest frequency.   
These galaxies all show excess noise at delay times of about a  day  that further increase for longer delays, as is typical of previously known AGN.  
However, from one Quarter to another the overall amplitude of the SF can change significantly.   
Thus the activity in these cases is not consistent with the simple auto-regressive or damped random walk models.  This was found by \citet{Kasliwal_etal15} to be the case for the more dominant AGN in Kepler studies. 

In most of the SFs, the power at the shortest delay times measured (an hour) is dominated by the Poisson noise of the target and sky and instrumental noise such as readout noise, and this explains the tendency to flatten out at delays of a day or less.  The brightest sources, though, clearly show physical activity registering all the way down to one hour.   In most of the cases, a knee occurs in the SF between 2 and 20 days where the LC nearly flattens to zero slope, indicating little additional variability at long timescales. 
 
The second most active, KIC 5686822, has SF power law index of $\sim 1.0$ around 1 - 6 day delay times, yet the index falls to nearly zero by 10 day timescales. LC excursions over $\sim 6$ days are typically at the 5 mmag level. 

Structure Functions react to periodic behavior by dropping to lower values at integer multiples of the period.  Several of the SFs consistently show clear and deep minima (KIC 10645722, KIC 5686822, KIC 7986325, and KIC 4148802), but  the amplitudes and delay times of the minima change over the observing.   An exception is KIC 7986325 which appears to maintain a single period but has big amplitude swings.  It is conceivable that this one is  contaminated by a variable star.   We are working on simulations to determine if this quasiperiodic behavior is likely to be statistical artifacts arising from power at longer frequencies (red leak), or from residuals of the trends in the Kepler data, or real quasiperiodic behavior in the accretion disks of BHs on the time scales of 20 - 40 days.
In the case they are Kepler artifacts, these galaxies remain AGN candidates since astrophysical activity is the best explanation for why these LCs could not be fully detrended by the cotrending analyses even though the amplitudes of variation are relatively small.

%and they are consistent with quasiperiodic behavior because

The conversion from SF to PSD is complex unless they are both simple power laws, therefore we have also created  PSDs (online data, http://www.astro.umd.edu/$\sim$eshaya/AGN/) from our LCs.  These also flatten at frequencies above $\nu$ = 1/day or so, and rise to lower frequencies, but most begin to flatten at ~10d$^{-1}$.  A table (also online) gives the parameters of fits ($\nu_{brk}, \alpha, \beta$) to the broken power law: 
\begin{eqnarray*}
PSD(\nu) &=& PSD(\nu_{brk}) \left(\frac{\nu}{\nu_{brk}}\right)^{-\alpha}~~  \mbox{for $0.1d^{-1} <  \nu < \nu_{brk}$} \nonumber \\
        &=& PSD(\nu_{brk}) \left(\frac{\nu}{\nu_{brk}}\right)^{-\beta}~~  \mbox{for  $\nu  > \nu_{brk}$} \nonumber 
\end{eqnarray*}
where $\nu_{brk}$ is the frequency of the break from the white noise regime to the target's variability signal.  At frequencies below 0.1d$^{-1}$, these PSDs are often not fit well by a power law.  
The fit parameter values vary widely in many of the candidates again indicating that the amplitude of the variations changes over time.  Of course, this would be expected if there is significant power on longer timescales.

\section{Conclusions}

AGN selection criteria have traditionally been based on imaging and spectroscopic techniques in the various wavelength bands (see \citet{Mushotzky_04} for a discussion). 
However when the light from the AGN is significantly less than that from the host galaxy the unique AGN spectral or imaging signatures can be difficult to recognize and result in severe incompleteness at even moderate redshift.  
Even in the x-ray band, where the contrast with stellar emission processes is the greatest, at low luminosities ($<10^{42}$ ergs/sec), AGN emission can be confused with the emission from ultra-luminous x-ray (ULX) sources, and at even lower luminosities with emission from bright x-ray binaries. 
The virtual ubiquity of optical variability from AGN combined with the very high signal to noise in the Kepler data open up a totally new channel for detecting AGN and allow direct comparison with other survey techniques. 

Using the Kepler Mission instrument, we measured unprecedented detailed light curves of $\sim$500 galaxies.
We were able to improve on the long term trending, in many cases, over the light curves generated by the project pipeline by using larger apertures and re-estimating optimal coefficients for the cotrending basis vector detrending step. 
Although non-optimal for the noise level of the 30 minute integrations, the larger apertures (typically $5 \times 5$ pixels) greatly reduced the longer timescale noise in the light curves by being less sensitive to PSF shape changes and centroid motions.   The project's derived cotrending basis vectors nicely removed most of the remaining instrumental trends.

Of the 474 galaxies with long baselines and imaged on quiet CCDs, 19 galaxies ($\sim$4\%) show excess activity indicative of having an active galactic nucleus.  Specifically, they had more variability at 15 days than at 5 hrs. 
The other roughly 460 galaxies were quiet, varying typically at levels below 5 milli-magnitudes over 15 days.
Many of our candidates have variability at levels that are not discernible by just viewing the LC, as our technique of using the SF ratio can detect quite low levels of variability.  
In the SFs of about half of the active galaxies, 10, show a knee at delay times of between 4 to 20 days above which the rate of rise in the amplitude of variability with delay times slows considerably.  But for four of these, the cause of the decrease toward longer time scales in the SF is due to periodicity.  It will require further analysis to determine if this is due to residual Kepler sensitivity trends, statistical fluctuations from breaking up the time series into short, roughly 90 day, sections, or quasiperiodic behavior of the AGNs.

We plan on obtaining optical spectra of the Kepler AGN candidates (some are already acquired) to determine whether they show spectral AGN signatures, but even that is not a certain measure, since a fair fraction of x-ray selected AGN  do not show either optical or near-IR AGN associated emission lines \citep{Smith_etal_14}.  Spectra could reveal if the narrow- or broad-line regions of these candidates are different in composition or structure from AGN with more pronounced variability.

The techniques developed in this paper have great applications to observations with LSST or the Zwicky Transient Factory which both should have sufficient precision, cadence and time coverage to reveal similar variability characteristics to what we have found with Kepler. We have a program to obtain more AGN with the K2 mission.  While the K2 observations consist of single 70 - 80 day time series for each field, the structure functions from Kepler show that this is sufficiently long to detect more candidates and obtain unique information on their power spectrum or structure functions.

\acknowledgments
The authors were supported by Kepler guest observing grants NNX12AC95G and NNX13AC27G from NASA.
All of the data presented in this paper were obtained from the Mikulski Archive for Space Telescopes (MAST).
STScI is operated by the Association of Universities for Research in Astronomy, Inc., under NASA contract
NAS5–26555. Support for MAST for non–HST data is provided by the NASA Office of Space Science via grant NNX13AC07G and by other grants and contracts.
This paper includes data collected by the Kepler mission. 
Funding for the Kepler mission is provided by the NASA Science Mission directorate. 
This research has made use of NASA's Astrophysics Data System and the NASA Exoplanet Archive, which is operated by the California Institute of Technology, under contract with NASA under the Exoplanet Exploration Program.
%\appendix

%\section{Appendicial material}

%10645722,5686822
\clearpage

%11716536,7986325
\clearpage
% 6751969,7691427
\clearpage
%12553112,11768473
\clearpage
% 2142191,12556836
\clearpage
%11808151, 5683305
\clearpage
% 8024526,5511084
\clearpage
%6714622,4148802
\clearpage
%10402746,9509125
\clearpage
% 8884097


\begin{thebibliography}{}
\bibitem[Barr \& Mushotzky(1986)]{BarrMushotzky_86}Barr, P. \& Mushotzky, R. F. 1986, \nat, 320, 421
\bibitem[Bershady et al.(1998)]{Bershady_etal_98}Bershady, M. A., Trevese, D., \& Kron, R. G. 1998, \apj, 496, 103
\bibitem[Borucki et al.(2010)]{Borucki_etal10}Borucki, Koch, Basri, et al. 2010, Science, 327, 977
\bibitem[Carini \& Ryle(2012)]{CariniRyle12}Carini \& Ryle. 2012, \apj, 749, 70
\bibitem[Chen \& Wang(2015)]{ChenWang15}Chen \& Wang. 2015, \apj, 805, 80
\bibitem[Cicio et al.(2014)]{Cicio_etal14}Cicio, D. et al 2014, http://arxiv.org/pdf/1412.1488v2.pdf, submitted to \aap
\bibitem[Crisiani et al.(1996)]{Cristiani_etal96}Cristiani, S., Trentini, S., La Franca, F., et al. 1996, \aap, 306, 395
\bibitem[Condon et al.(1998)]{Condon_etal98}Condon, J. J. et al. 1998, \aj 115, 1693 (1998).
\bibitem[Edelson \& Malkan(2012)]{EdelsonMalkan2012}Edelson, R. \& Malkan, M. 2012, \apj, 751, 52
\bibitem[Edelson et al.(2014)]{Edelson_etal2014}Edelson, R., Vaughan, S., Malkan, M., Kelly, B. C., Smith, K. L., Boyd, P. T.; \& Mushotzky, R. 2014, \apj, 795,2 
\bibitem[Faber \& Dressler(1977)]{FaberDressler77}Faber, S. M. \& Dressler, A. 1977, \aj 82, 187
\bibitem[Frank, King \& Raine(2002)]{FrankKingRaine02}Frank J., King A., \& Raine D.J., Cambridge University Press, 2002, Accretion Power in Astrophysics: Third Edition
\bibitem[Jarrett et al.(2000)]{Jarret_etal00}Jarrett, T. H., Chester, T., Cutri, R., et al. 2000, \aj, 119, 2498
\bibitem[Hook et al.(1994)]{Hook_etal94}Hook, I. M.,McMahon, R. G., Boyle, B. J. \& Irwin, M. J. 1994,\mnras, 268, 305
\bibitem[Kasliwal et al.(2015)]{Kasliwal_etal15}Kasliwal, V. P., Michael S. V. \& Gordon T. R. 2015. \mnras, 451, 4328
\bibitem[Kawaguchi et al.(1998)]{Kawaguchi_etal98}Kawaguchi, Mineshige, Umemura \& Turner. 1998, \apj, 504, 671
\bibitem[Klesman \& Sarajedini(2007)]{KlesmanSarajedini_07}Klesman, A. \& Sarajedini, V. 2007, \apj, 665, 225
\bibitem[Kolodziejczak et al.(2010)]{Kolo10}Kolodziejczak, J. J., Caldwell, D. A., van Cleve, J. E., et al., in Proc. SPIE 2010, 7742, 38
\bibitem[MacLeod et al.(2010)]{MacLeod_etal10}MacLeod, C. L., Ivezi\'c, \v{Z}., Kochanek, C. S., et al. 2010, \apj, 721, 1014
\bibitem[Mushotzky(2004)]{Mushotzky_04}Mushotzky, R. F. in Supermassive Black Holes in the Distant Universe, Astrophysics and Space Science v. 308 [editor: Barger, A. J.], 2004 (astro-ph/0405144) 
\bibitem[Quintana et al.(2010)]{Quin10}Quintana, E. V., Jenkins, J. M., Clarke, B. D., et al., in Proc. SPIE 2010, 7740, 64
\bibitem[Revalski et al.(2014)]{Revalski_etal14}Revalski, M. et al. 2014, \apj, 785, 60
\bibitem[Sarajedini et al.(2011)]{Sarajedini_etal11}Sarajedini, V. L., Koo, D. C., Klesman, A. J., et al. 2011, \apj, 731, 97
\bibitem[Schechter(1976)]{Schechter76}Schechter, P. 1976, \apj, 203, 297
\bibitem[Shakura \& Sunyaev(1973)]{ShakuraSunyaev73}Shakura \& Sunyaev. 1973, \aa, 24, 337
\bibitem[Smith et al.(2014)]{Smith_etal_14} Smith, K. L., Koss, M., Mushotzky, R. F. 2014, \apj,  794, 112 
\bibitem[Stumpe et al.(2012)]{Stumpe12}Stumpe, M. C., Smith, J. C., Van Cleve, J. E., et al. 2012, \pasp, 124, 985
\bibitem[Trevese et al.(2008)]{Trevese_etal_08}Trevese, D., Boutsia, K., Vagnetti, F., Cappellaro, E., \& Puccetti, S. 2008, \aap,488, 73
\bibitem[Ulrich et al.(1997)]{Ulrich_etal1997}Ulrich, Maraschi \& Urry. 1997, \araa, 35, 445–502
\bibitem[Villforth et al.(2010)]{Villforth_etal10}Villforth, C., Koekemoer, A. M., \& Grogin, N. A. 2010, \apj, 723, 737
\bibitem[Webb \& Malkan(2000)]{WebbMalkan_00}Webb, W. \& Malkan, M. 2000, \apj, 540, 65
\bibitem[Mushotzky et al.(2011)]{Mushotzky_etal11}Mushotzky, Edelson, Baumgartner \& Gandhi. 2011, \apjl, 743, L12
\bibitem[Wehrle et al.(2013)]{Wehrle_etal13}Wehrle, Wiita, Unwin, et al. 2013, \apj, 773, 89

\end{thebibliography}
\end{document}